\documentclass[11pt, epsfig]{article}
\usepackage{epsfig, amsmath, amssymb, amsthm, times}
\usepackage{graphicx}
\DeclareMathOperator{\esssup}{ess\,sup}

\newtheorem {thm}{Theorem}[section]

\theoremstyle{defintion}
\newtheorem {df}[thm]{Definition}

\theoremstyle{remark}
\newtheorem{rem}[thm]{Remark}

\theoremstyle{example}
\newtheorem{ex}[thm]{Example}

\theoremstyle{assumption}

\def\pf{{\it Proof.\;}}

\def\P{{\mathbb P~}}
\def\R{{\mathbb R}}
\def\N{{\mathbb N}}

\def\Z{{\mathbb Z}}
\def\SS{{\mathbb S}}

\def\lbl{\label}
\def\be{\begin{equation}}
\def\ee{\end{equation}}
\def\p{\partial}
\def\qed{\square}

\def\t{\mathsf{T}}

\def\one{\mathbf{1}}

\title{The nonlinear heat equation on dense graphs and graph limits
}
\author{Georgi S. Medvedev 
\thanks{
Department of Mathematics, Drexel University, 3141 Chestnut Street,
Philadelphia, PA 19104;
{\tt medvedev@drexel.edu}
}
}

\begin{document}
\maketitle
\begin{abstract} 
The continuum limit of coupled dynamical systems is an approximate
procedure, by which the dynamical problem on a sequence of  large graphs is
replaced by an evolution integral equation on a continuous spatial 
domain. While this method has been widely used in the analysis of
pattern formation in nonlocally coupled networks, its mathematical
basis remained little understood.

In this paper, we use the combination of ideas and results from
the theory of graph limits and nonlinear evolution equations
to provide a rigorous mathematical justification for 
taking the continuum limit and to extend this method to cover 
many complex networks, for which it has not been applied
before.  Specifically, for dynamical networks on convergent sequences of simple
and weighted graphs, we prove convergence of solutions of the
initial-value problems for discrete models to those of the limiting 
continuous equations. In addition, for sequences of simple graphs converging to 
\{0, 1\}-valued graphons, it is shown that the convergence rate depends 
on the fractal dimension of the boundary of the support of the graph limit.   
These results are then used to study the regions of continuity of chimera states 
and the attractors of the nonlocal Kuramoto equation on certain multipartite graphs.
Furthermore, the analytical tools developed in this 
work are used in the rigorous  justification of the continuum limit for networks on
random graphs that we undertake in a companion paper \cite{Med13a}.

As a by-product of the analysis of the continuum limit on deterministic and random
graphs, we identify the link between this problem and the convergence analysis of several
classical numerical schemes: the collocation,  Galerkin, and Monte-Carlo methods.
Therefore, our results can be used to characterize convergence of these approximate
methods of solving initial-value problems for nonlinear evolution equations with 
nonlocal interactions.
\end{abstract}

\section{Introduction}\lbl{sec.intro}
Coupled dynamical systems on graphs represent many diverse models
throughout the natural sciences and technology. Examples range from
regulatory and neuronal
networks in biology \cite{LaiCho01,BenHan97, MZ12, Swi80},
to Josephson junctions and coupled lasers in physics \cite{LiErn92, PhiZan93, WatStr94}, 
to communication, sensor, and power networks in technology \cite{DorBul12, Med12},
to name a few. Compared to partial differential equations and lattice dynamical
systems, the analysis of networks meets a new principal challenge: the rich variety
and possible complexity of the underlying graphs. The algebraic methods of graph
theory \cite{Biggs, Chung-Spectral} have been useful in understanding the contribution
of the network topology to certain aspects of networks dynamics, especially in problems
involving synchronization \cite{Med12, MZ12}. The continuum limit of nonlocally coupled
dynamical networks is one of few analytical approaches that have a potential for elucidating
dynamics of a broad class of networks \cite{KurBat02, AbrStr06, WilStr06, 
 GirHas12, OmeWol12,OmeRie12}.
In this limit, the solutions of the 
initial value problems (IVPs) for evolution equations on large discrete 
networks are approximated by those for the limiting  integro-differential
equations posed on continuous spatial domains. This limiting
procedure has been used to study the mechanisms 
of some very interesting effects such as chimera states 
\cite{KurBat02,AbrStr06}, 
multistability \cite{WilStr06,GirHas12},
synchronization, and the coherence-incoherence transition \cite{OmeWol12}.
However, a rigorous justification for taking the continuum limit in
nonlocally models was lacking. In this paper, we use the combination of 
techniques from the theory of evolution equations \cite{EvaPDE} and the recent theory
of graph limits \cite{BorChay06, BorChay08, LovSze06, LovSze07, LovGraphLim12}
 to provide such justification for a large class of dynamical 
models on deterministic graphs. In fact, some of the tools that we develop 
in this work come in useful in the analysis of the  continuum limit of dynamical systems 
on random graphs undertaken in a companion paper \cite{Med13a}.

To motivate the forthcoming analysis of the continuum limit
in the nonlocally coupled systems, we first review several representative 
examples. In \cite{WilStr06}, Wiley, Strogatz, and Girvan 
studied a nonlocally coupled system of phase oscillators
\be\lbl{Kuramoto}
\dot\phi_i=\omega+{1\over n} \sum_{j=i-k}^{i+k} \sin\left(\phi_j-\phi_i\right),
\ee
where $\phi_i:~\R^+\to \SS^1:=\R/2\pi\Z,\; i\in [n]:=\{1,2,3,\dots,n\}$ is interpreted as the phase
of oscillator~$i$, $\omega$ is the intrinsic frequency, and the sum models the interactions between
oscillator~$i$ and $k$ of its nearest neighbors from each side
(cf. \cite{Kur84, KurBat02}). The oscillators are located on a ring and indexed
by integers from $\Z/n\Z$. By recasting (\ref{Kuramoto}) in uniformly rotating
frame of reference, one can absorb $\omega$. Thus, below we set $\omega=0$.

\begin{figure}
\begin{center}
{\bf a}\hspace{0.1 cm}\includegraphics[height=1.8in,width=2.0in]{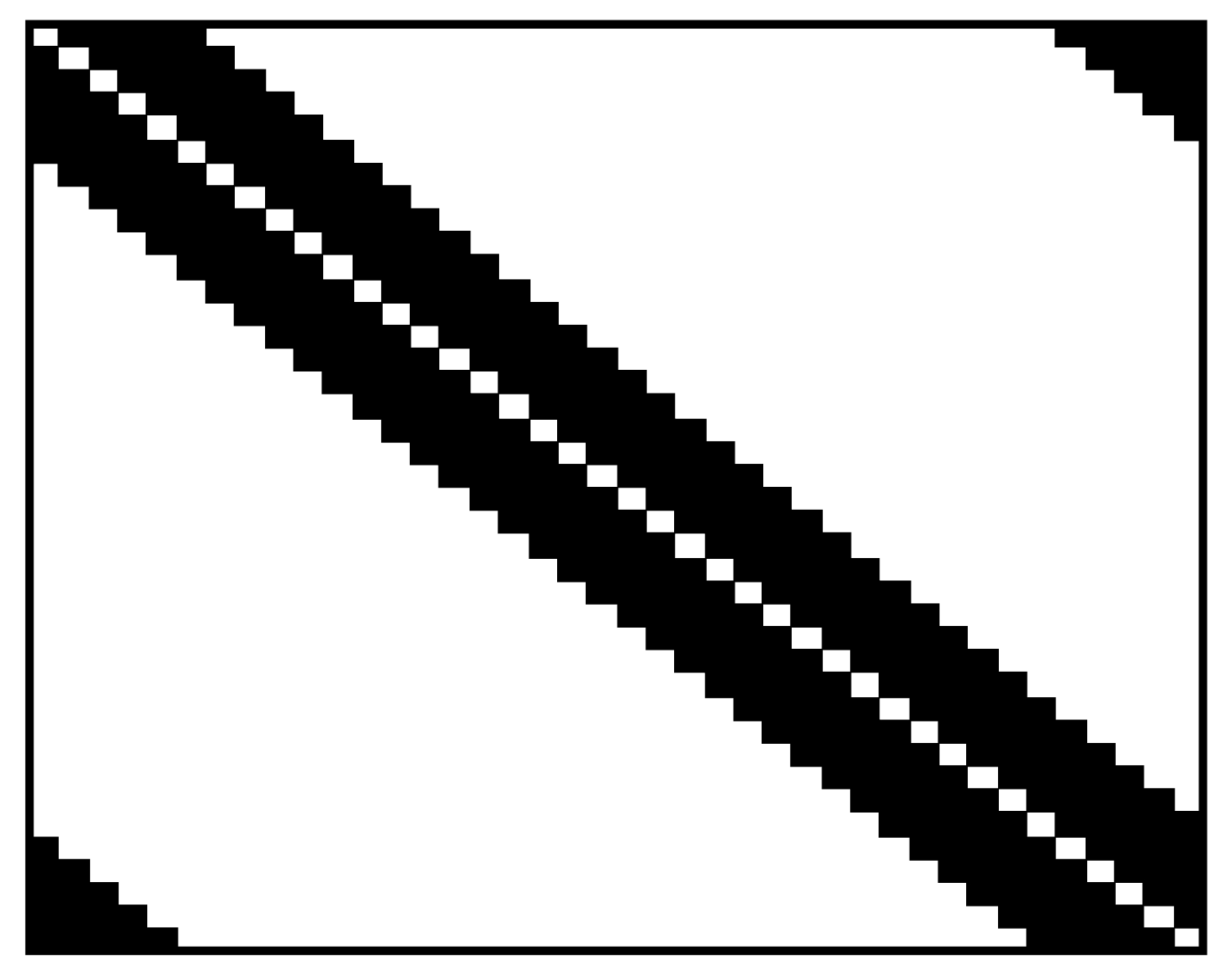}\hspace{1.0cm}
{\bf b}\hspace{0.1 cm}\includegraphics[height=1.8in,width=2.0in]{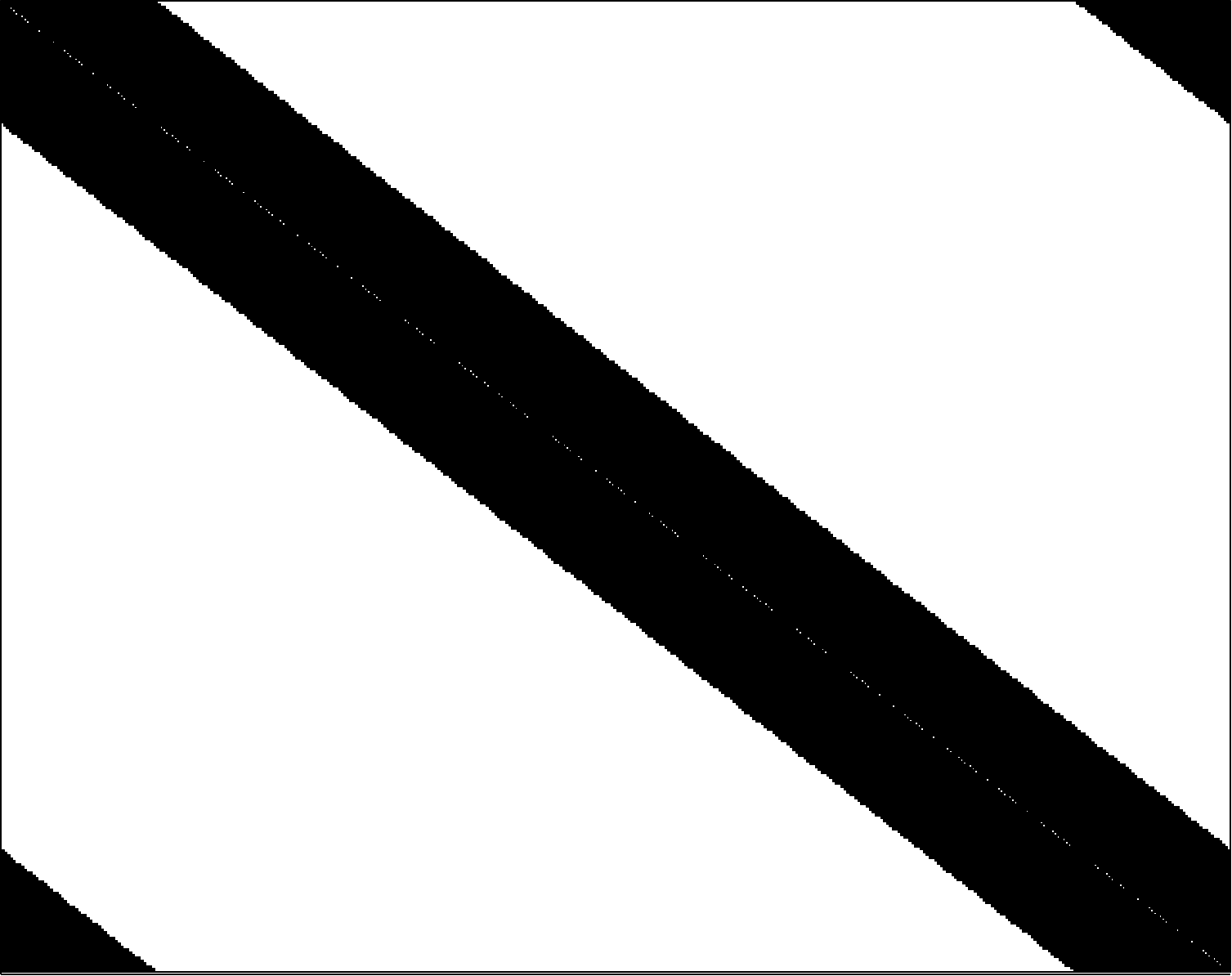}
\end{center}
\caption{The plot of the support of the  function $W_{G_n}$ representing the adjacency matrix
of the $k$-nearest-neighbor graph $G_n$ ({\bf a}) and that of its limit $W_G$ ({\bf b}).
}
\lbl{f.1}
\end{figure}

It is instructive to view (\ref{Kuramoto}) as a system of differential
equations on graph $G_n=\langle V(G_n), E(G_n)\rangle$ with the 
vertex set $V(G_n)=[n]$ and the edge set 
$$
E(G_n)=\left\{ (i,j)\in [n]^2:~ 0<\mbox{dist}(i,j)\le k\right\},\;\mbox{where}\; 
\mbox{dist}(i,j) =\min\{ |i-j|, n-|i-j|\}.
$$
Let $W_{G_n}:~I^2\to\{0,1\}$ such that 
$$
W_{G_n}(x,y)=1\;\mbox{if}\; (i,j)\in E(G_n)\;\mbox{and}\; (x,y)\in [(i-1)n^{-1}, in^{-1})\times [(j-1)n^{-1}, jn^{-1}).
$$
Here and below, $I$ denotes $[0,1]$, the spatial domain of the
continuum limits considered in this paper.
The plot of the support of $W_{G_n}(x,y)$ in Fig.~\ref{f.1}a provides the  pixel picture 
of the adjacency matrix of $G_n$ \cite{Biggs}.  In Fig.~\ref{f.1}a and
in similar plots throughout this paper, 
we place the origin of the unit square  in the top left corner of the plot to emphasize 
the correspondence between $W_{G_n}$ and the adjacency matrix of $G_n$. 
As $n\to \infty$, $\{W_{G_n}\}$ converges to the $\{0,1\}$-valued function $W_G(x,y)$,
whose support is shown in Fig.~\ref{f.1}b.

In \cite{WilStr06}, the analysis of the attractors  of (\ref{Kuramoto}) employs  the continuum limit
of (\ref{Kuramoto}). 
%
 Specifically, let $k=rn$ for some fixed $r\in (0,1]$. After interpretting the right 
hand-side of (\ref{Kuramoto}) as a Riemann sum and sending $n\to\infty$, 
in the uniformly rotating frame of coordinates  (\ref{Kuramoto}) formally becomes
\be\lbl{cont}
{\p\over \p t}\phi(x,t)=\int_I W_G(x,y)
\sin\left(\phi(y,t)-\phi(x,t)\right)dy,
\ee
where $\phi(x,t)$ describes the evolution of the continuum of oscillators
distributed over  $I$. 
Equation (\ref{cont}) is called the continuum (thermodynamic) limit of
(\ref{Kuramoto}).\footnote{There is another form of the continuum limit
for the Kuramoto model \cite{StrMir91, Str2000, OttAnt08, Lai09}.
It is formulated in terms of the density characterizing the state of the 
continuous system. We do not consider this limit in the present paper.}
  
The continuum equation (\ref{cont}) has a family of steady state 
solutions
\be\lbl{twist}
\theta^{(q)}(x,t)=2\pi qx+c, \; q\in \Z, \; c\in \R,
\ee
called $q-$twisted states. In \cite{WilStr06}, the stability analysis of
the continuous twisted states (\ref{twist}) was used to study their
discrete counterparts, which are the steady state solutions of 
(\ref{Kuramoto}) ($\omega=0$) for finite $n$. The stability analysis 
in \cite{WilStr06} can, in fact, be 
completely translated into the discrete setting. However, suppose we replace
the family of $k$-nearest-neighbor graphs in (\ref{Kuramoto})
by a family of small-world graphs (see Fig.~\ref{ff.1}a). Then not only
does the continuum limit provide a convenient setting for the stability
analysis but also the twisted states, as the steady states 
of the Kuramoto model, exist only in the limit 
as the number of oscillators goes to infinity (see  Fig.~\ref{ff.1}b)~\cite{Med13b}. 
Therefore, in this case the continuum
limit affords the analysis of the asymptotic behavior of solutions 
of the Kuramoto model for large $n$, which is not otherwise feasible in the discrete
setting. The Kuramoto-Battogtokh model generating chimera states \cite{KurBat02}
is another example, where the contnuum limit seems to be critical
for  understanding the nontrivial dynamics in the discrete systems.
We will return to the discussion of chimera states in Section \ref{sec.chimera}.

\begin{figure}
\begin{center}
{\bf a}\hspace{0.1 cm}\includegraphics[height=1.8in,width=2.0in]{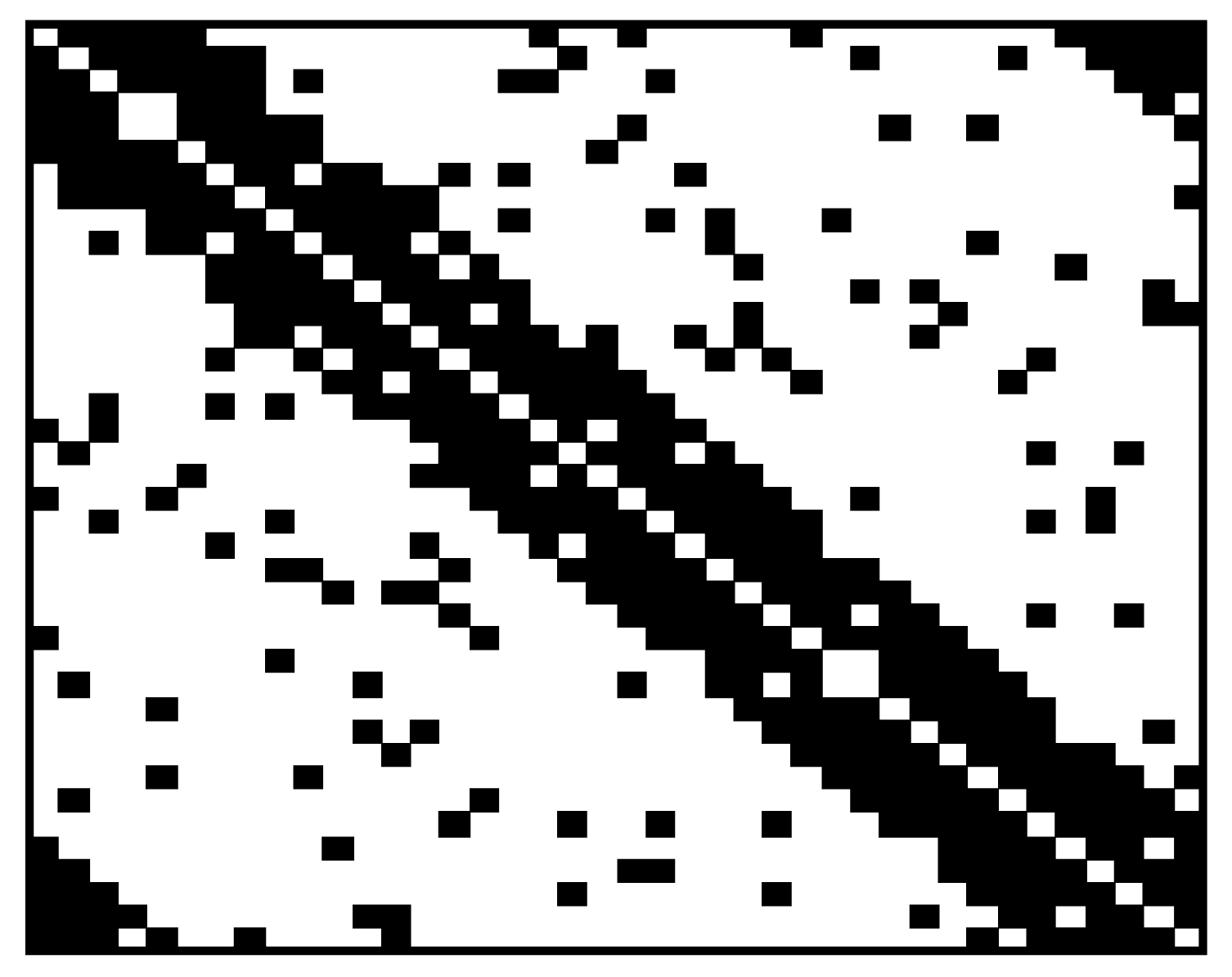}\hspace{1.0cm}
{\bf b}\hspace{0.1 cm}\includegraphics[height=1.8in,width=2.0in]{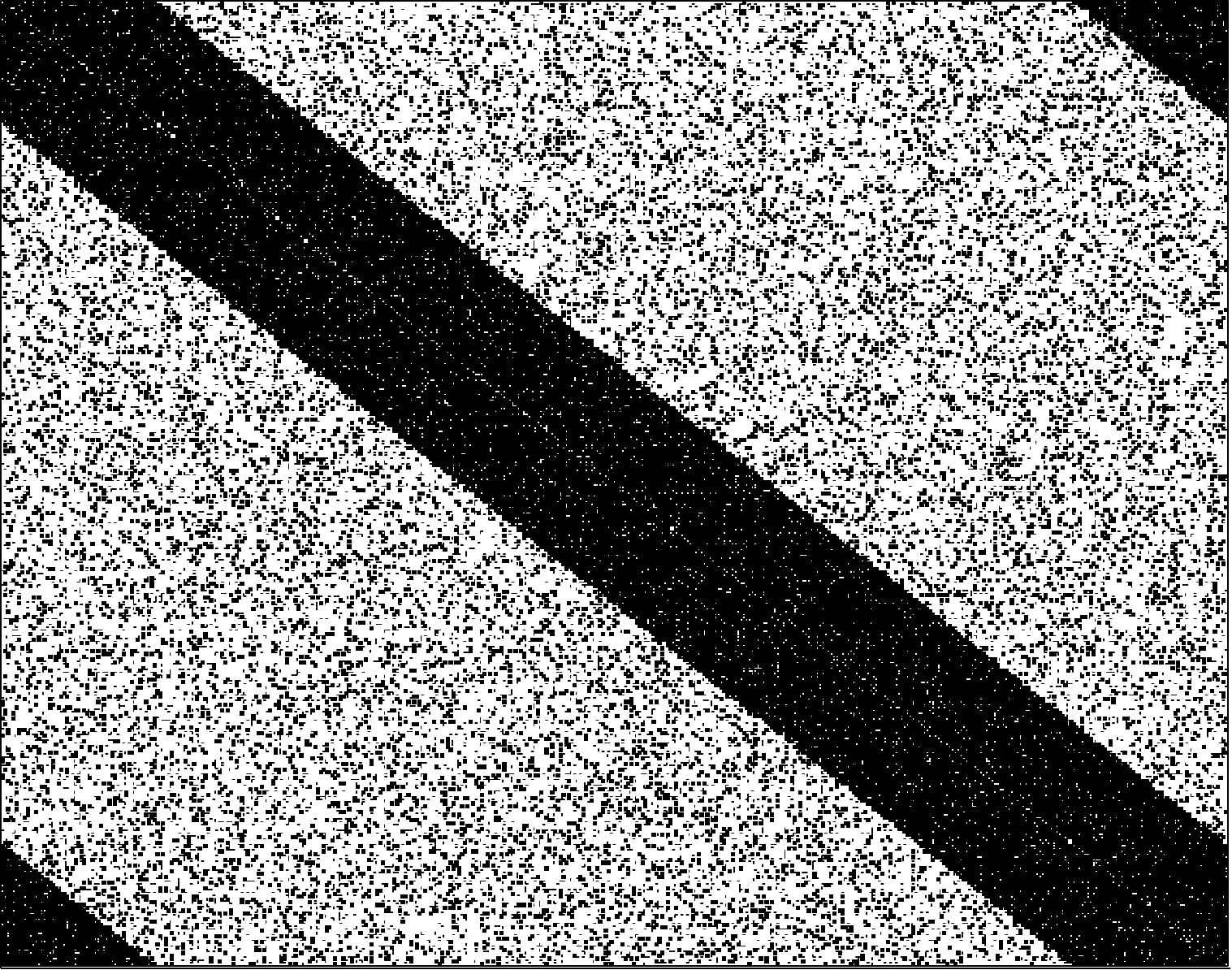}
\end{center}
\caption{ {\bf a}) The pixel picture of a small-world graph obtained from that shown in 
Fig.~\ref{f.1}{\bf a}
by replacing a random set of the local connections by randomly chosen long-range ones.
{\bf b}) The pixel picture for a large small-world graph. 
}
\lbl{ff.1}
\end{figure}

These examples lead to the following questions.
\begin{description}
\item[(A)] Does the continuum model (\ref{cont})  truly approximate the dynamics of the discrete
model (\ref{Kuramoto}) for large finite $n$? If so, in what sense the solutions of the integro-differential
equation
approximate  those of (\ref{Kuramoto}) with $\omega=0$?
\item[(B)] How big is the class of network topologies for which one can use the continuum
limit? Is it restricted to the special graphs like $k-$nearest-neighbor one on a ring?
Can it be applied, for instance, to the small world networks, the original motivation for the
analysis in \cite{WilStr06}?
\end{description}

The function $W_G$ shown in Fig.~\ref{f.1}b is the limit of the functions
$\{W_{G_n}\}$ (Fig.~\ref{f.1}a) representing the adjacency matrices of the
$k$-nearest neighbor family of graphs $\{G_n\}$. The latter  is an example
of a convergent graph sequence and $W_G$ is the corresponding graph
limit \cite{LovGraphLim12}. We will explain the meaning of the limit of
a graph sequence in Section~\ref{sec.limit}. Meanwhile, we refer to the 
geometric interpretation of the adjacency matrix  for the $k$-nearest-neighbor
graph in Fig.~\ref{f.1}a, which suggests the limiting pattern of $\{W_{G_n}\}$
as $n\to\infty$ (see Fig.~\ref{f.1}b). Likewise, the pixel picture of the large
small-world graph in Fig.~\ref{ff.1}b suggests the (piecewise constant)
limit for the small-world family of graphs, which in turn can be used in the
derivation of the continuum model like (\ref{cont})  \cite{Med13b}.
These observations hint on the possible relevance of the theory of graph limits for constructing
the continuum limits for dynamical networks. We explore this 
relation for dynamical systems on convergent families of deterministic graphs
in this paper and extend this approach to random networks in \cite{Med13a}.
Interestingly, in the process of justifying the continuum limit, we discovered the
link between this problem and that of convergence of several classical
numerical methods. Specifically, we show that dynamical networks on simple 
and weighted graphs analyzed in Sections~\ref{sec.simple} and \ref{sec.weight}
can be interpreted as the discretizatizations of the continuum evolution equation
by the collocation method and the Galerkin method respectively. Furthermore,
the analysis of the continuum limit for networks on random graphs in \cite{Med13a} 
features a similar connection with the Monte-Carlo method. Therefore, in addition to the  
rigorous justification of taking the continuum limit for a large class of dynamical
networks, our results characterize convergence of these numerical methods
for solving IVPs for certain nonlinear integro-differential equations.

This paper is organized as follows.
We review the necessary background on graph limits in Section~\ref{sec.limit}.
In Section~\ref{sec.formulate}, we discuss the heat equation on graphs
and graph limits. Here, we extend a classical linear heat equation
on graphs  to allow nonlinear diffusion. This extension covers 
many dynamical networks
arising in applications including  coupled oscillator models like (\ref{Kuramoto}).
In the same section, we formally define the continuum limit for 
dynamical networks of a convergent sequence of dense (weighted) 
graphs. In this limit, the discrete diffusion operator becomes
an integral operator with the kernel representing the limit of
the infinite family of graphs. We show that the IVP for the
limiting equation is well-posed and admits a unique solution in
$C^1(\R;L^\infty(I))$. Further, in Theorem~\ref{thm.reg}, we specify 
assumptions on the kernel and the initial conditions, which guarantee 
that the solutions of the IVPs remain continuous in space over 
subdomains of $I$.
This result is used to characterize the attractors of the
continuum model. In particular, we apply it to study 
the regions of continuity of the chimera states and attractors
of the Kuramoto equation on certain multipartite graphs (see Section~\ref{sec.examples}).
The rest of the paper is focused on  studying the relation between
the solutions of the IVPs for discrete networks and  and their continuum counterparts.
In Section~\ref{sec.simple}, for sequences 
of  simple graphs converging to $\{0,1\}-$valued graphons, we  show that
the rate of convergence depends on the fractal dimension 
of the boundary of the support of the graph limit.
This shows explicitly how the geometry of the graphon affects
the accuracy of the continuum limit. In Section~\ref{sec.weight},
we analyze networks on convergent 
weighted graph sequences. 
 The results of this paper
are illustrated with the discussion of the dynamics of two concrete
models: the Kuramoto-Battogtokh nonlocal system generating 
chimera states \cite{KurBat02} and the Kuramoto equation on the
half and complete bipartite graphs (cf.~Section~\ref{sec.examples}). 
The final section, Section~\ref{sec.conclusion}, 
contains concluding remarks.

\section{Graph limits}\lbl{sec.limit}
\setcounter{equation}{0}
 
In this section, we review several definitions and results from the theory of graph 
limits that we will need later. In our brief tour through graph limits, we mainly follow 
\cite{BorChay11} and \cite{Pikh10}. For the full exposition of this powerful theory 
with many diverse applications, we refer an interested reader to the pioneering 
papers by Lov{\' a}sz and Szegedy  \cite{LovSze06, LovSze07}, and Borgs, Chayes, 
Lov{\' a}sz, S{\'o}s, and Vesztergombi \cite{BorChay06, BorChay08}; and to the
monograph \cite{LovGraphLim12}.

An undirected graph $G=\langle V(G), E(G)\rangle$ 
without loops and multiple edges is called simple. $V(G)$ stands for the 
set of nodes and $E(G)\subset V(G)\times V(G)$ denotes the edge set. 
\begin{figure}
\begin{center}
{\bf a}\hspace{0.1 cm}\includegraphics[height=1.8in,width=2.0in]{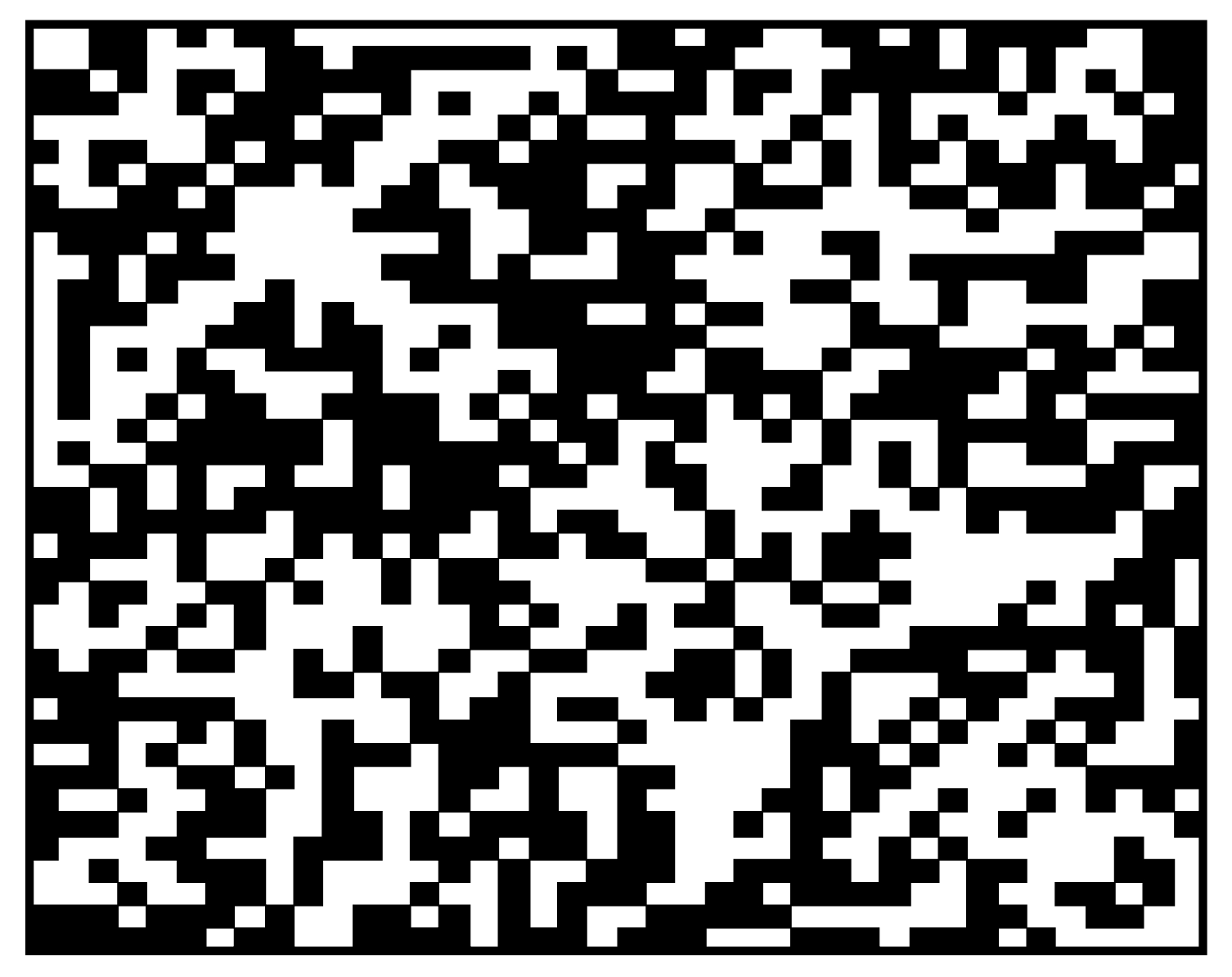}\hspace{1.0cm}
{\bf b}\hspace{0.1 cm}\includegraphics[height=1.8in,width=2.0in]{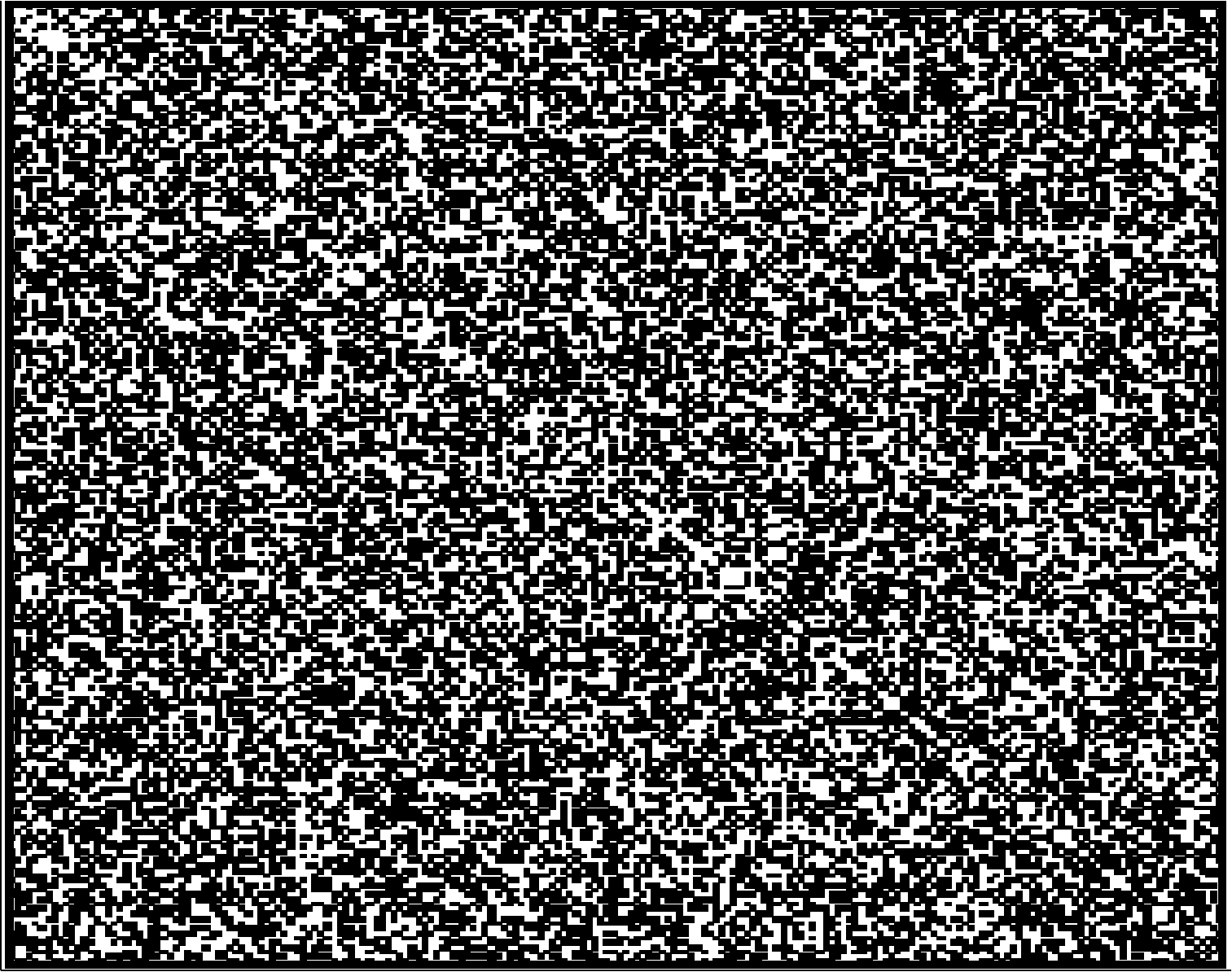}
\end{center}
\caption{ {\bf a}) The pixel picture of the Erd\H{o}s-R\'{e}nyi graph  $G(40,0.5)$.
The edge between a pair of distinct nodes is inserted with probability $0.5$. 
{\bf b}) The pixel picture of $G(600,05)$.
}
\lbl{ff.2}
\end{figure}

Let $G_n=\langle V(G_n), E(G_n)\rangle, n\in\mathbb{N}$ be a sequence of dense 
(simple) graphs, i.e., $|E(G_n)|=O(|V(G_n)|^2)$, where $\left|\cdot\right|$ denotes the cardinality of a set.
The convergence of the graph sequence  $\{G_n\}$ is defined in terms of the homomorphism densities
\be\lbl{hdense}
t(F,G_n)={\mbox{hom}(F,G_n)\over \left|V(G_n)\right|^{|V(F)|}}.
\ee
Here, $F=\langle V(F), E(F)\rangle$ is a simple graph and $\mbox{hom}(F,G_n)$ stands for
 the number of homomorphisms (i.e., adjacency preserving maps
$V(F)\to V(G_n)$).
In probabilistic terms,  (\ref{hdense}) is the likelihood of a random
map $h:~V(F)\to V(G_n)$ to be a homomorphism. 

\begin{df}\lbl{df.convergent}\cite{LovSze06, BorChay08}
The sequence of graphs $\{G_n\}$
is called convergent if $t(F,G_n)$ is convergent for every simple graph 
$F$.\footnote{In the theory of graph limits, convergence in Definition~\ref{df.convergent}
is called left-convergence. Since this is the only convergence of graph sequences  used 
in this paper, we refer to the left-convergent sequences as convergent.} 
\end{df}

It turns out that the limiting object can be represented by a measurable symmetric function
$W: I^2\to I$. We recall that $I$ stands for $[0,1]$. 
Such functions are called graphons. The set of all graphons is 
denoted by $\mathcal{W}_0$.

\begin{thm}\cite{LovSze06}
For every convergent sequence of simple graphs, there is  $W\in\mathcal{W}_0$
such that
\be\lbl{t-to-t}
t(F,G_n)\to t(F,W):=\int_{I^{|V(F)|}} \prod_{(i,j)\in E(F)} W(x_i,x_j) dx 
\ee
for every simple graph $F$.
Moreover, for every $W\in\mathcal{W}_0$ there is a sequence of graphs 
$\{G_n\}$ satisfying 
(\ref{t-to-t}).
\end{thm}

The cut-norm is important for describing  the metric  properties of graphons. 
For any integrable function and, in particular, for any graphon $W\in\mathcal{W}_0$,  
$$
\|W\|_\qed =\sup_{S,T\in \mathcal{L}_I} \left| \int_{S\times T} W(x,y) dxdy\right|
$$
is called the cut-norm of $W$. Here, $\mathcal{L}_I$ stands for the set of all Lebesgue measurable
subsets of  $I$. The cut-distance between two graphons $W$ and $U$ 
is defined by 
$$
\delta_\qed(U,W)=\inf_{\phi}\|U-W^\phi\|_\qed,\;\;
$$
where $W^\phi(x,y):=W(\phi (x),\phi (y))$ and $\phi$ ranges over all measure-preserving 
bijections of $I$.   The infinum over all $\phi$ is used to make the cut-distance between graphons
invariant with respect to graph isomorphisms, as well as some other transformations that do not change
the asymptotic properties of the graph sequences  (see \cite{BorChay08, LovGraphLim12} for 
more details). 
A graph sequence is convergent if and only if it is Cauchy in the cut-distance \cite{BorChay08}.

Graph limits are the equivalence classes of graphons
$$
[W]=\left\{U\in\mathcal{W}_0:~\delta_\qed(U,W)=0\right\}.
$$
With a customary abuse of notation, we refer to both $W$ and $[W]$ as graphons.
The pseudo-metric $\delta_\qed(\cdot,\cdot)$ induces the metric on 
$\chi=\{[W]:~W\in\mathcal{W}_0\}$.
The metric space $(\chi,\delta_\qed)$ is compact \cite{LovSze07}.

\begin{figure}
\begin{center}
{\bf a}\hspace{0.1 cm}\includegraphics[height=1.8in,width=2.0in]{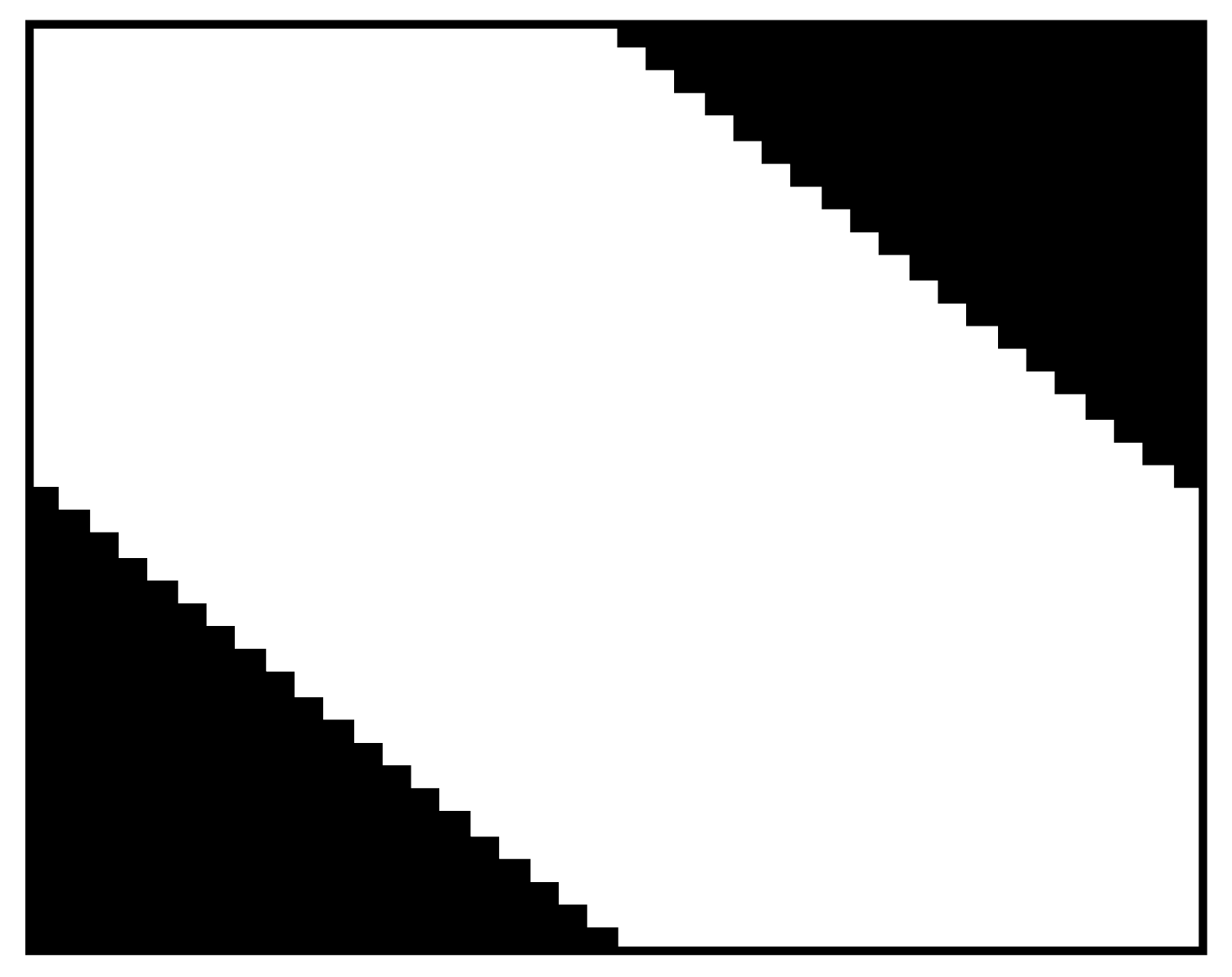}\hspace{1.0cm}
{\bf b}\hspace{0.1 cm}\includegraphics[height=1.8in,width=2.0in]{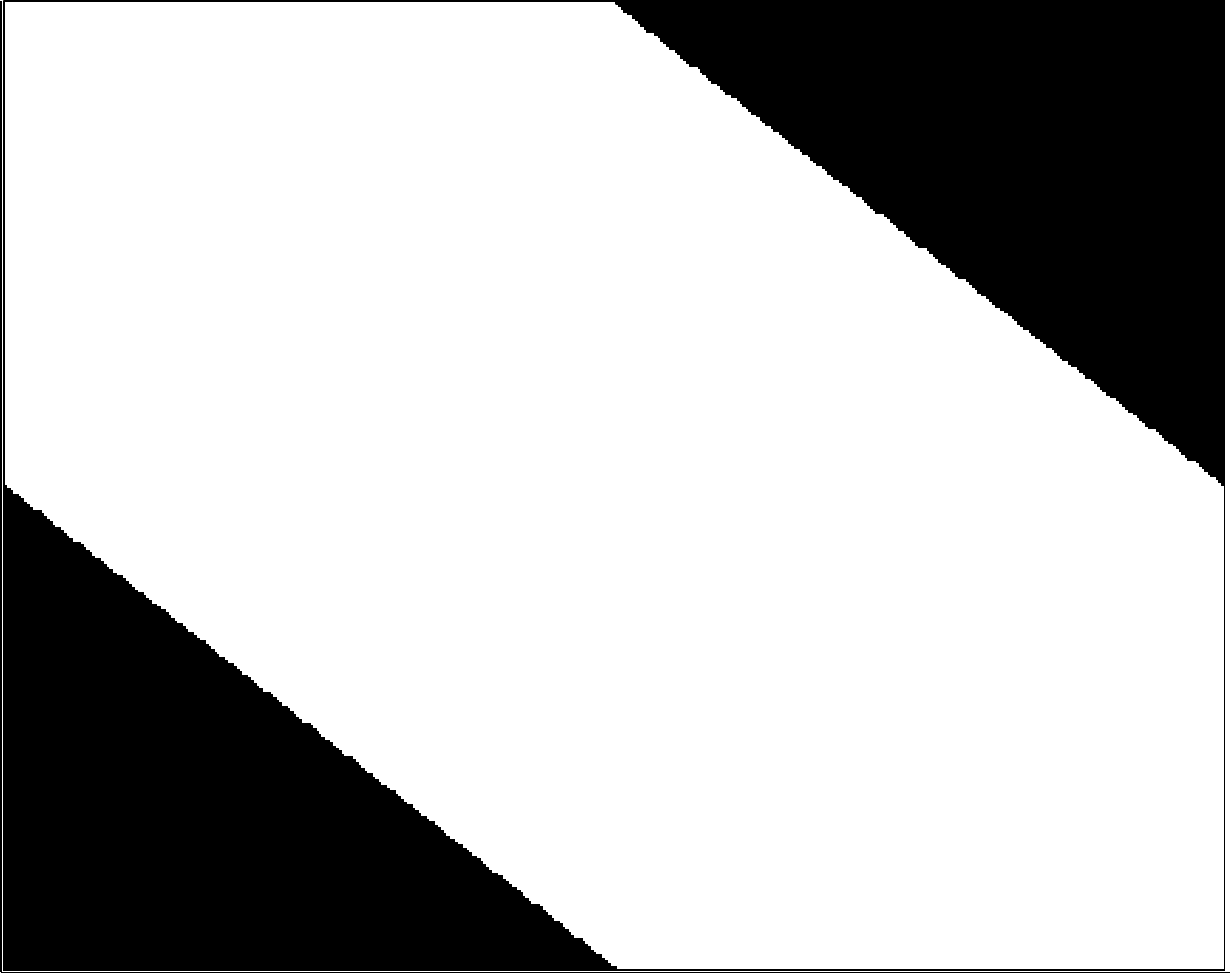}
\end{center}
\caption{ {\bf a}) The pixel picture of the half-graph $H_{20,20}$.
{\bf b}) The limit of $\{W_{H_{n,n}}\}$. 
}
\lbl{ff.3}
\end{figure}

We now describe a simple construction of convergent graph sequences that will be used 
in the analysis of the continuum limit of dynamical networks below. 
Consider a sequence of simple graphs $\{G_n\}$ on $n$ nodes. Define
\be\lbl{pixel}
W_{G_n}(x,y)=\left\{ \begin{array}{ll} 1, & \mbox{if}\; (i,j)\in E(G_n)\;\mbox{and}\;
(x,y)\in \left[{i-1\over n}, {i\over n}\right)\times \left[{j-1\over n}, {j\over n}\right),\\
0,& \mbox{otherwise}.
\end{array}\right.
\ee
The support of $W_{G_n}$ provides  the pixel picture of the adjacency
matrix of $G_n$ (see Fig.~\ref{ff.1}a), and  $[W_{G_n}]$ is the corresponding graphon.
Note that $[W_G]$ is invariant under relabeling the nodes of $G$
while $W_{G_n}$ is not.  
The graph sequence $\{G_n\}$ is convergent if $W_{G_n}$ converge 
with respect to the cut-norm. In particular, since for any integrable
function $W\in\mathcal{W}_0$
$$
\|W\|_\qed\le \|W\|_{L^1(I^2)},
$$
convergence of $\{W_{G_n}\}$ in the $L^1$-norm implies convergence
of the graph sequence $\{G_n\}$. The deterministic networks 
analyzed in this paper are actually convergent with respect to the stronger
$L^1$-norm. However, the convergence of graphons with 
respect to the cut-norm does not in general imply that with respect to
$L^1$-norm. For instance, the sequence of Erd\H{o}s-R{\'e}nyi
graphs with edge density $p\in (0,1)$ is convergent to the constant 
function $p$ on $I^2$, $\mbox{Const}(p)$ \cite{LovSze06, BorChay08}, 
while no sequence of $\{0,1\}$-valued graphons
can converge to $\mbox{Const}(p)$ with $p\in (0,1)$ in the $L^1$-norm.  
In particular, $L^1$-estimates for graphons are insufficient for the
analysis of the continuum limits of
networks on random graphs \cite{Med13a}.

We conclude this section we several examples of convergent graph sequences.
\begin{ex}\lbl{ex.ER}\cite{LovSze06, BorChay08} 
The Erd\H{o}s-R\'{e}nyi graphs. 
Let $p\in (0,1)$ and consider a sequence of random graphs 
$G(n,p)=\langle V(G(n,p)),  E(G(n,p))\rangle$,
$V(G(n,p))=[n]$ such that the probability
$\P\{(i,j)\in E(G(n,p))\}=p$ for any $(i,j)\in [n]^2$ (see Fig.~\ref{ff.2}a). 
Then for any simple graph $F,$ $t(F, G(n,p))$ is convergent with probability
$1$ to $p^{|E(F)|}$ as $n\to\infty$ \cite{BorChay08}. 
Thus, $\{G(n,p)\}$  is a convergent sequence with the limit given by the 
constant graphon $p$.  The pixel picture of $W_{G(n,p)}$ in Fig.~\ref{ff.2}b
provides the intuition behind the graph limit for $\{G(n,p)\}$. Note that 
for large $n$, the plot of the support of $W_{G(n,p)}$ resembles
that of the constant function if looked at from a distance.
In fact, the limiting graphon reflects  the asymptotic density 
of connections in $G(n,p)$ as  $n\to\infty$. Using the strong law of large numbers, 
one can show that $\|W_{G(n,p)} - p\|_\square\to 0$  as $n\to\infty$  with probability
$1$. Thus,  $\{W_{G(n,p)}\}$ is convergent in the cut-norm but not in the $L^1$-norm. 
\end{ex}


\begin{ex}\lbl{ex.half}\cite{LovSze06} The half-graphs. Let $H_{n,n}=\langle V(H_{n,n}), E(H_{n,n})\rangle$ 
be a bipartite graph on $2n$ nodes such that
$$
V(H_{n,n})=\{1,2,\dots,n,1^\prime, 2^\prime,\dots, n^\prime \},\;
E(H_{n,n})=\{( i,j^\prime) \in V(H_{n,n})\times V(H_{n,n}):~ i\le j\}
$$
(see Fig.~\ref{ff.3}a).
The sequence $\{ H_{n,n}\}$ converges to the graphon $[H]$ where
$H: I^2\to I$ is the characteristic function of the set
$\{ (x,y):~ |x-y| \ge 1/2\}$ (see Fig.~\ref{ff.3}a).In this example, $\{W_{H_{n,n}}\}$ 
converges  to $H$ pointwise, and, by the dominated convergence theorem, 
in the $L^1$-norm.
\end{ex}

\section{The formulation of the problem} \lbl{sec.formulate}
\setcounter{equation}{0}
\subsection{The heat equation on discrete and continuous domains}
Let $G_n=\langle V(G_n), E(G_n), W(G_n)\rangle$ be  a sequence of weighted graphs,
where $V(G_n)=[n]$ and $E(G_n)$ are the sets of nodes and edges respectively;
and $W(G_n)\!:\ [n]^2\to[-1, 1]$ is a symmetric weight matrix of the form
$$
(W(G_n))_{ij}=\left\{\begin{aligned}&w_{ij}^{(n)}, && (i, j)\in E(G_n),\\ &0, 
&&{\rm otherwise.}\end{aligned}\right.
$$
If $G_n$ is a  simple  graph, $W(G_n)$ is a $\{0,1\}$-valued matrix.

By  the nonlinear heat equation on $G_n$ we mean  the system of
differential equations
\be\lbl{dheat}
{d\over dt} u_i^{(n)}(t)=\lambda_i^{(n)}\sum_{j=1}^n w_{ij}^{(n)} 
D\left(u^{(n)}_j-u^{(n)}_i\right),\; i\in [n],
\ee
where $u^{(n)}(t)=\left( u^{(n)}_1(t),u^{(n)}_2(t),\dots,u^{(n)}_n(t)\right)^\t$,
and $\lambda_i^{(n)}$ are scaling coefficients. 
The function $D:~\R\to\R$ is  Lipschitz continuous 
\be\lbl{Lip}
\left|D(u)-D(v)\right|\le L |u-v| \;\forall u,v\in \R.
\ee
Throughout this paper, we will 
use $\lambda_i^{(n)}=n^{-1}$. However, other scalings may also be used.

\begin{rem}\label{rem.generalize}
Our analysis applies to a more general class of equations 
\be\lbl{more-general}
{d\over dt} u_i^{(n)}(t)=\lambda_i^{(n)}\sum_{j=1}^n w_{ij}^{(n)} 
D\left(u^{(n)}_j-u^{(n)}_i\right) + f_i(t,u^{(n)}),\; i\in [n],
\ee
where functions $f_i(t,u), i\in [n],$ can be taken, for instance, to be
continuous in $t$ and Lipschitz continuous in $u$:
$$
\left|f_i(t,u)-f_i(t,v)\right|\le L |u-v| \;\forall u,v,t \in \R, \; i\in [n].
$$
To keep the presentation simple, we will restrict the  analysis to the case of 
(\ref{dheat}). 
It is straightforward to extend our results to cover (\ref{more-general}).
\end{rem}

If $D(u)=u$, the coupling operator on the right-hand side
of (\ref{dheat}) is the graph Laplacian, and  Equation (\ref{dheat})
becomes the linear heat equation on $G_n$. The linear heat equation
has many applications in combinatorial problems such as random walks on graphs
\cite{Chung-Spectral}, and dynamical problems, e.g., analysis of consensus 
protocols \cite{Med12}. In this paper, we focus on the nonlinear heat
equation, which provides the framework for a large class of dynamical networks.
In particular, the  Kuramoto equation (\ref{Kuramoto}) is of this type.

In the remainder of this paper,  we will derive and justify
the continuum counterpart of (\ref{dheat})
\be\lbl{nheat}
{\p\over\p t} u(x,t) =\int_{I} W(x,y) D\left( u(y,t)-u(x,t)\right) dy.
\ee
The kernel $W$ will be specified separately for each class of problems that we consider
below.
 
\subsection{The well-posedness of the IVP}
Before setting out to study the relation between solutions of the discrete and continuous
heat equations (\ref{dheat}) and (\ref{nheat}), we first address the well-posedness 
of the IVP for (\ref{nheat}).

It is  convenient to interpret the solution of the IVP for (\ref{nheat}),
$u(x,t)$, as a vector-valued map $\mathbf{u}:[0,T]\to L^\infty(I)$. Throughout this paper,
we will use the bold font to denote the vector-valued function $\mathbf{u}(t)$
corresponding to a function of two variables $u(x,t)$.

\begin{thm}\lbl{thm.wellposed}
Suppose $D$ is Lipschitz continuous, $W\in L^\infty(I^2)$, and $\mathbf{g}\in L^\infty(I)$. 
Then for any $T>0$, there exists a unique solution of the IVP for (\ref{nheat}) 
$\mathbf{u}\in C^1(\R;L^\infty(I))$ subject to the initial condition $\mathbf{u}(0)=\mathbf{g}$.
\end{thm}
\pf\; The proof of Theorem~\ref{thm.wellposed} is based on the contraction mapping principle
(cf.~\cite[Theorem~1.1, Chapter VII]{DalKre70}). We include the details for completeness.

Rewrite the IVP for (\ref{nheat}) as the integral equation
\be\lbl{fix}
\mathbf{u}=K\mathbf{u},
\ee
where
$$
[K\mathbf{u}](x,t):=\mathbf{g} +\int_0^t \int_I W(x,y) D\left( u(y,s)-u(x,s)\right)dyds.
$$
Let $M_\mathbf{g}$ be a metric subspace of $C(0,\tau; L^\infty (I))$ (where $\tau>0$ will be
specified later) consisting of functions $\mathbf{u}$ satisfying $\mathbf{u}(0)=\mathbf{g}$.
Then (\ref{fix})  is the fixed point equation for the operator $K: M_{\mathbf{g}}\rightarrow M_{\mathbf{g}}$. 
We show below that $K$ is a contraction for a small $\tau>0$.

Indeed, let 
\be\lbl{tau}
\tau\le (4L\|W\|_{L^\infty(I^2)})^{-1},
\ee
where $L$ is the Lipschitz constant of $D(\cdot)$. For any $\mathbf{u}, \mathbf{v}\in M_\mathbf{g}$
we have
$$
\left\| K\mathbf{u} - K\mathbf{v} \right\|_{M_\mathbf{g}} = \max_{t\in [0,\tau]}
\left\| K\mathbf{u}(t) - K\mathbf{v}(t) \right\|_{L^\infty (I)}
$$
$$
\le \max_{t\in [0,\tau]} \mathop{\esssup}_{x\in I}
 \int_{I\times [0,t]}   \left| W(x,y)\right| 
\left|  D\left( u(y,t)-u(x,t)\right)- D\left( v(y,t)-v(x,t)\right) 
\right| dy dt 
$$
$$
\le \tau L \|W\|_{L^\infty(I^2)} \max_{t\in [0,\tau]}  
\left\{  \int_I \left|u(y,t)-v(y,t)\right| dy+ 
\left\| \mathbf{u}(t)-\mathbf{v}(t)\right\|_{L^\infty (I)} \right\}
$$
$$
\le
2\tau L \|W\|_{L^\infty(I^2)}\max_{t\in [0,\tau]} \left\|\mathbf{u}(t)-\mathbf{v}(t)\right\|_{L^\infty (I)}.
$$
Thus, by (\ref{tau}) we have
\be\lbl{contract}
\| K\mathbf{u}-K\mathbf{v} \|_{M_\mathbf{g}} \le {1\over 2} 
\| \mathbf{u}-\mathbf{v} \|_{M_\mathbf{g}} .
\ee

By the Banach contraction mapping principle, there exists a unique
solution of the IVP for (\ref{nheat}) $\bar{\mathbf{u}}\in M_\mathbf{g}\subset C(0,\tau; L^\infty(I))$.
Using $\bar{\mathbf{u}}(\tau)$ as the initial condition, the local solution
can be extended  to $[0, 2\tau]$, and, by repeating this argument, 
to $[0, T]$ for any $T>0$. In a similar fashion, we can prove the existence and
uniqueness of the solution of the IVP for (\ref{nheat}) on $[-T,0]$ for any
$T>0$.
Furthermore, since the integrand
in (\ref{fix}) is continuous as a map 
$L^\infty(I)\to L^\infty(I)$, $\mathbf{u}$ is continuously differentiable. 
Thus, we have a classical solution
of the IVP for (\ref{nheat}) on the whole real axis.\\
$\qed$


\subsection{Spatial regularity}
The classical heat equation, as  a parabolic partial differential equation, has a strong 
smoothening property. 
Regardless of the regularity of the initial data, the solution of the IVP 
for the classical heat equation is a smooth function of the space variables
for all positive times. No such mechanism is present in the heat
equation on graph limits. Below we show that
the spatial regularity of solutions of the IVP is  determined by the 
regularity of graphon $W$ and initial condition $\mathbf{u}(0)$.

\begin{thm}\lbl{thm.reg}
Let $D:\R\to\R$ be a Lipschitz continuous function and 
$J=(\alpha,\beta)\subset I$. Suppose 
for all $x\in J$ and for almost all $y\in I$,
$W\in  L^\infty(I^2)$ has a weak derivative ${\p\over\p x}W(x,y)$
and 
\be\lbl{weak-assumption}
\esssup_{y\in I} \left\|{\p\over\p x}W(\cdot,y)\right\|_{L^2(J)}\le C_1,
\ee
for some $C_1>0$.
Then for any $0<T<\infty$, all $t\in [0,T],$ and 
$\alpha<\alpha^\prime<\beta^\prime<\beta,$ the solution of the 
IVP for (\ref{nheat}) satisfies\footnote{$H^1(J)$ stands
for the Sobolev space of all Lebesgue measurable functions
$f$ on an open interval $J\subset \R^1$ such that $f$ and
its distributional derivative $f_x$ are in $L^2(J)$ \cite{CheMil12}.}
$$
\mathbf{u}(t)\in H^1(J^\prime),\; J^\prime=(\alpha^\prime, \beta^\prime),
$$
provided $\mathbf{u}(0)\in L^\infty(I)\cap H^1(J)$.
\end{thm}
\pf\;
Let $T>0$ be arbitrary but fixed, and 
$$
h_0={1\over 2} \min\{ \alpha^\prime-\alpha, \beta-\beta^\prime\}.
$$
Then  for $0<h<h_0$, the difference quotient
$$
\xi(x,t)={u(x+h,t)-u(x,t)\over h}
$$
is a well-defined function on $\Omega_T=J^\prime\times [0,T]$. Further,
for $(x,t)\in \Omega_T$, $\xi(x,t)$ 
satisfies the following equation
\begin{eqnarray}\nonumber
{\p \over \p t}\xi(x,t)&=&
\int_I W(x,y) h^{-1}\left\{ D\left(u(y,t)-u(x+h,t)\right)-
D\left(u(y,t)-u(x,t)\right)\right\}dy\\
\lbl{xiode}
&+&\int_I D^h_x W(x,y) D\left(u(y,t)-u(x+h,t)\right) dy,
\end{eqnarray}
where
$$
D^h_x W(x,y)={W(x+h,y) -W(x,y)\over h}.
$$
By multiplying both sides of (\ref{xiode}) by $\xi(x,t)$ and
integrating both sides of the resultant equation over $J^\prime$ with respect
to $x$, we have
\begin{eqnarray}\nonumber
{1\over 2}\int_{J^\prime} {\p \over \p t}\xi(x,t)^2 dx &=&\int_{J^\prime \times I}
 W(x,y) h^{-1}\left\{ D\left(u(y,t)-u(x+h,t)\right)-
D\left(u(y,t)-u(x,t)\right)\right\}\xi(x,t) dxdy\\
\nonumber
&+&\int_{J^\prime\times I} 
D^h_x W(x,y) D\left(u(y,t)-u(x+h,t)\right) \xi(x,t) dxdy\\
\lbl{re-xiode}
&=:& T_1+T_2.
\end{eqnarray}
Using $\mathbf{u}\in C(0,T;L^\infty(I))$, Lipschitz continuity of $D(\cdot)$,
and the triangle inequality, we have
\be\lbl{boundD}
\max_{t\in [0,T]} \esssup_{(x,y)\in I^2} |D(u(y,t)-u(x,t))|\le 2L
\|\mathbf{u}\|_{C(0,T; L^\infty(I))}=:C_2.
\ee
Furthermore,  using Fubini's theorem, (\ref{weak-assumption}), and 
the standard results  for the difference quotients
(see, e.g., Theorem~5.8.3~\cite{EvaPDE}), we have 
\be\lbl{boundDW}
\|D_x^h W\|_{L^2(J^\prime \times I)} \le \esssup_{y\in I}\|D_x^h W\|_{L^2(J^\prime)} 
 \le C_3 \esssup_{y\in I} \|{\p\over \p x} W(\cdot,y)\|_{L^2(J)} \le C_4,
\ee
and, likewise,
\be\lbl{bound-xi0}
\|{\boldsymbol\xi}(0)\|_{L^2(J^\prime)} \le C_5 \| \mathbf{u}(0)\|_{H^1(J)},
\ee 
where positive constants $C_4$ and $C_5$ are independent of $h\in(0,h_0)$.

Using (\ref{Lip}), we bound the first term on
the right hand side (\ref{re-xiode})
\be\lbl{T1}
|T_1|\le  \|W\|_{L^\infty(I^2)} \int_{J^\prime\times I} L \xi(x,t)^2 dxdy= L \|W\|_{L^\infty(I^2)}\|{\boldsymbol\xi}(t)\|_{L^2(J^\prime)}^2.
\ee
For the second term, we use (\ref{boundD}), (\ref{boundDW}),
and the Cauchy-Schwarz inequality
\begin{eqnarray}\nonumber
|T_2| &\le &  C_2  \int_{J^\prime\times I}
\left| D^h_x W  {\boldsymbol\xi}(t)  \right| dxdy
 \le C_2  \|D_x^h W \|_{L^2(J^\prime \times I)}
\|{\boldsymbol\xi}(t)\|_{L^2(J^\prime)}  \\
\lbl{T2}
&\le & C_2 C_4 \|{\boldsymbol\xi}(t)\|_{L^2(J^\prime)}.
\end{eqnarray}

By combining (\ref{re-xiode}), (\ref{T1}), and (\ref{T2}), we have
$$
{d\over dt} \|{\boldsymbol\xi}(t)\|^2_{L^2(J^\prime)} \le C_6\|{\boldsymbol\xi}(t)\|^2_{L^2(J^\prime)}+C_7, \; 
C_6=2L \|W\|_{L^\infty(I^2)} +C_7, \; C_7=C_2 C_4,
$$
where inequality $2\|{\boldsymbol\xi}(t)\|_{L^2(J^\prime)} \le
\|{\boldsymbol\xi}(t)\|^2_{L^2(J^\prime)}+1$ was used.
Using  Gronwall's inequality, we obtain
\begin{eqnarray}\nonumber
\|{\boldsymbol\xi}(t)\|^2_{L^2(J^\prime)} &\le&
\left(\|{\boldsymbol\xi}(0)\|^2_{L^2(J^\prime)} +{ C_7 \over C_6}\right)
\exp\{C_6T\}\\
\lbl{bound-xi}
&\le& 
\left( C_5^2 \| \mathbf{u}(0)\|^2_{H^1(J)}      +{ C_7 \over C_6}\right)
\exp\{ C_6T\}, \; t\in[0,T].
\end{eqnarray}
The last inequality  yields a uniform in $h\in (0,h_0]$ bound on the difference quotient
$\|{\boldsymbol\xi}(t)\|_{L^2(J^\prime)}$.
Using the properties of the difference quotients 
(cf. Theorem~5.8.3~\cite{EvaPDE}), we conclude  that $\mathbf{u}(t)\in H^1(J^\prime)$ 
for all  $t\in (0,T]$.\\
$\qed$

\section{Networks on simple graphs}\lbl{sec.simple}
\setcounter{equation}{0}
In this and in the following sections, we prove that the solution
of the IVP for appropriately chosen continuous problem (\ref{nheat})
approximates the solutions of the discrete problems (\ref{dheat})
when $n$ is sufficiently large. We prove this result for two classes
of convergent graph sequences. In this section, we consider
the case of a sequence of simple graphs converging to a
$\{0,1\}$-valued graphon, and we study a more general case
of convergent sequences of weighted 
graphs\footnote{For weighted graphs, one can also define 
convergence by extending the notion of the homomorphism density
for this case (see \cite{LovSze06} for details). We do not discuss this 
generalization here, because for the problems that we study in this 
paper a simpler (and stronger) form of convergence, convergence
in $L^1-$norm, is sufficient (see Section~5).}
in the next section. 
We single out networks on $\{0,1\}$-valued graphons for two reasons. 
First, many coupled oscillator models  fit into 
this framework (see, e.g.,  \cite{WilStr06, GirHas12} and \S\ref{sec.half}). 
Second, for this class of networks we can explicitly estimate the accuracy
of approximation of the solutions of the discrete models by those 
of their continuum limits in terms of the network size and the geometry
of the graphon of the network (cf. Theorem~\ref{coloc-rate}).
This result is important, because it reveals the structural properties of the 
graphs shaping the accuracy of the thermodynamic limit.

Let $W:I^2\to\{0,1\}$ be a symmetric measurable function.
We denote the support of $W$ by
$$
W^+=\{ (x,y)\in I^2:\; W(x,y)\neq 0\}
$$
and its boundary by $\p W^+$.

For  convenience, we rewrite the IVP for (\ref{nheat})
\begin{eqnarray}\lbl{sheat}
{\p\over\p t} u(t,x)&=&\int_I W(x,y) D\left( u(y,t)-u(x,t)\right) dy,\\
\lbl{sheat-ic}
u(x,0) &=& g(x).
\end{eqnarray}
Throughout this section, to simplify presentation we assume that
$g(x)$ 
is a step function.

Next, we define a sequence of discrete problems. 
To this end, we fix $n\in\N$,  divide $I$ into $n$ subintervals
\be\lbl{partition}
 I^{(n)}_1=\left[0, {1\over n}\right),\; I^{(n)}_2=\left[{1\over n},{2\over n}\right), \dots,
I_n^{(n)}=\left[ {n-1\over n},1 \right),
\ee
and define a sequence  of simple graphs $G_n=\langle V(G_n), E(G_n) \rangle$ 
such that
$V(G_n)=[n]$ and
$$
E(G_n)=\{(i,j)\in [n]^2:\; (I_i^{(n)}\times I_j^{(n)})\cap W^+\neq\emptyset \}.
$$

The IVP for the nonlinear heat equation on $\{G_n\}$, a discrete
counterpart of (\ref{sheat}),
is given by
\begin{eqnarray}\lbl{dsheat}
{d\over dt} u_i^{(n)}(t) &=& n^{-1} \sum_{j: (i,j)\in E(G_n)} D(u^{(n)}_j-u^{(n)}_i),\\
\lbl{dsheat-ic}
u^{(n)}_i(0)&=&g^{(n)}_i,\; i\in [n].
\end{eqnarray}
There are many ways of approximating $g(x)$ by $g_n(x)$. For concreteness,
we assign $g_i^{(n)}$ the average value of $g(x)$ on $I_i$: 
\be\lbl{mean-g}
g_i^{(n)}=n \int_{I^{(n)}_i} g(x) dx.
\ee

To compare the solutions of the discrete and continuous models, it is
convenient to represent the discrete function $u^{(n)}=(u^{(n)}_1,
u^{(n)}_2,\dots, u^{(n)}_n)^\t$ as a step function on $I$ as follows
\be\lbl{step-fun}
u_n(x,t) = u^{(n)}_i, \;\mbox{if} \; x\in I^{(n)}_i.
\ee
Then $u_n(x,t)$ satisfies the following IVP
\begin{eqnarray}\lbl{coloc}
{\p\over\p t} u_n(t,x)&=&\int_I \hat W_n(x,y) D\left( u_n(y,t)-u_n(x,t)\right) dy,\\
\lbl{coloc-ic}
u(x,0) &=& g_n(x),
\end{eqnarray}
where 
$$
g_n(x)=g_i^{(n)} \;\mbox{if}\;  x\in I^{(n)}_i, i\in[n].
$$
and $\hat W_n (x,y)$ is the step function such that for $(x,y)\in
I^{(n)}_i\times I^{(n)}_j,$ 
$(i,j)\in [n]^2,$
\be\lbl{hat-W}
\hat W_n (x,y)= \left\{\begin{array}{ll} 1, & \mbox{if}\;  (I^{(n)}_i\times I^{(n)}_j)\cap W^+\neq\emptyset,\\
                                     0, & \mbox{otherwise}.
\end{array}
\right.
\ee

\begin{thm}\lbl{coloc-rate} 
Let $\mathbf{u}$ and $\mathbf{u_n}$ denote the vector-valued functions
corresponding to the solutions 
of (\ref{sheat}), (\ref{sheat-ic}), and (\ref{coloc})-(\ref{mean-g})
respectively.  Denote 
the upper box-counting
dimension of $\p W^+$ 
by $2b=\overline{\dim}_B \p W^+$
(cf. \S~3.1,\cite{Falc-Fractal}) and suppose
that $b\in [0.5, 1)$.
Then for any $\epsilon> 0$ and all sufficiently large $n$
\be\lbl{rate}
\| \mathbf{u} -\mathbf{u_n}\|_{C(0,T;L^2(I))} \le C_1
n^{-(1-b-\epsilon)},
\ee
where constant $C_1$ is independent of $n$.
\end{thm} 
\pf\; 
Denote $\xi_n(x,t)= u_n(x,t)-u(x,t)$. By subtracting
(\ref{sheat}) from (\ref{coloc}), we have
\begin{eqnarray}\nonumber
{\p \xi_n\over \p t} &=& \int_I \hat W_n(x,y)\left\{ D\left(u_n(y,t)-u_n(x,t)\right)-
D\left(u(y,t)-u(x,t)\right)\right\} dy\\
&+& 
\lbl{subtract1}
\int_I \left( \hat W_n(x,y)- W(x,y)\right) D\left(u(y,t)-u(x,t)\right)dy.
\end{eqnarray}
Next, we multiply both sides of (\ref{subtract1}) by $\xi_n(x,t)$ and
integrate over $I$
\begin{eqnarray}\nonumber
{1\over 2} \int_I {\p\over \p t} \xi_n(x,t)^2 dx &=& \int_{I^2} \hat W_n(x,y)\left\{ D\left(u_n(y,t)-u_n(x,t)\right)-
D\left(u(y,t)-u(x,t)\right)\right\}  \xi_n(x,t) dxdy\\
&+& 
\lbl{dxi2a}
\int_{I^2} \left(\hat W_n(x,y)-W(x,y)\right) D\left(u(y,t)-u(x,t)\right)\xi_n(x,t) dxdy.
\end{eqnarray}
Using the Lipschitz continuity of $D(\cdot)$,  $\|\hat W\|_{L^\infty (I^2)}=1$, 
the triangle inequality, and the Cauchy-Schwarz inequality,
we estimate the first term 
on the right-hand side of (\ref{dxi2a})
\begin{eqnarray}\nonumber
\left|
\int_{I^2} \hat W_n(x,y)\left\{ D\left(u_n(y,t)-u_n(x,t)\right)-
D\left(u(y,t)-u(x,t)\right)\right\}  \xi_n(x,t) dxdy
\right|
\\
\lbl{first}
\le L \int_{I\times I} \left|\left(\xi_n(y,t)-\xi_n(x,t)\right)
  \xi_n(x,t)\right| dxdy \le
2L\|{\boldsymbol\xi}_n(t)\|_{L^2 (I^2)}^2.
\end{eqnarray}

We estimate 
the second term on the right-hand side of  (\ref{dxi2a}),
using the Cauchy-Schwarz inequality and the bound on $D(\cdot)$ (cf. (\ref{boundD}))
\begin{eqnarray}\nonumber
\left|\int_{I^2} \left( \hat W_n(x,y)-W (x,y)\right) D\left(u(y,t)-u(x,t)\right)\xi_n(x,t) dxdy\right|\\
\nonumber
\le \esssup_{(x,y,t)\in I^2\times [0,T]} \left| D\left(u(y,t)-u(x,t)\right)\right| 
\left|\int_{I^2} \left( \hat W_n(x,y)-W (x,y)\right)\xi_n(x,t) dxdy\right|\\
\lbl{second}
\le C_2 \|W-\hat W_n\|_{L^2(I^2)} \|{\boldsymbol\xi}_n\|_{L^2(I)}
\end{eqnarray}
for some constant $C_2>0$ independent of $n$.

Using (\ref{first}) and (\ref{second}), from  (\ref{dxi2a}) we have
\begin{equation}\lbl{insert-1}
{d\over dt} \|{\boldsymbol\xi}_n\|^2_{L^2(I)} \le 4L \|{\boldsymbol\xi}_n\|^2_{L^2(I)}+
2C_2 \|W-\hat W_n\|_{L^2(I^2)}\|{\boldsymbol\xi}_n\|_{L^2(I)}.
\end{equation}

Let $\varepsilon>0$ be arbitrary but fixed, and set
$$
\phi_\varepsilon(t)=\sqrt{\|{\boldsymbol\xi}_n\|^2_{L^2(I)}+\varepsilon}.
$$
By (\ref{insert-1}),
\be\lbl{insert-2}
{d\over dt} \phi_\varepsilon(t)^2\le 4L \phi_\varepsilon(t)^2+
2C_2 \|W-\hat W_n\|_{L^2(I^2)}\| \phi_\varepsilon(t).
\ee
Since $\phi_\varepsilon(t)$ is positive on $[0,T]$, from
(\ref{insert-2}), we have
$$
{d\over dt} \phi_\varepsilon(t)\le 2L \phi_\varepsilon(t)+
C_2 \|W-\hat W_n\|_{L^2(I^2)}, \; t\in [0,T].
$$
By Gronwall's inequality,
\be\lbl{insert-3}
\sup_{t\in [0,T]} \phi_\varepsilon(t) \le \left( \phi_\varepsilon(0)
+{C_2\|W-\hat W_n\|_{L^2(I^2)}\over 2L}\right) \exp\{2LT\}.
\ee
Since $\varepsilon>0$ is arbirtrary, (\ref{insert-3}) implies
\be\lbl{Gron}
\sup_{t\in [0,T]}\|{\boldsymbol\xi}_n(t)\|_{L^2(I)}\le \left( \|\mathbf{g}-\mathbf{g}_n\|_{L^2(I)}
+{C_2\|W-\hat W_n\|_{L^2(I^2)}\over 2L}\right) \exp\{2LT\}.
\ee 

It remains to estimate $\|W-\hat W_n\|_{L^2(I^2)}$. 
To this end, consider the set of discrete cells $I_i^{(n)} \times I_j^{(n)}$ that covers the boundary
of the support of $W$
$$
J(n)=\{(i,j)\in[n]^2:~ (I_i^{(n)}\times I_j^{(n)})\cap \p W^+\neq\emptyset \} \;\mbox{and}\; 
C(n)=\left|J(n)\right|.
$$
Using one of several equivalent definitions of the upper box-counting dimension of 
a subset of $\R^n$, we have
$$
2b:=\overline{\dim}_B \p W^+=
\overline{\lim_{\delta\to 0}} {\log N_\delta (\p W^+)\over -\log\delta},
$$
where $N_\delta (\p W^+)$ is the number of cells of a $(\delta\times\delta)$-mesh  that intersect $\p W^+$
(see Equation (3.12)(iv) in \cite{Falc-Fractal}).
Thus, for any $\epsilon>0$ and all sufficiently large $n$, we have
$$
C(n)\le  n^{2(b+\epsilon)}.
$$

Since $W$ and $\hat W_n$ coincide on all cells $I_i^{(n)}\times I_j^{(n)}$ for which
$(i,j)\notin J(n)$, for any $\epsilon>0$ and all sufficiently large $n$, we have
\be\lbl{estimW}
\|W-\hat W_n\|_{L^2(I^2)}^2=\int_{I^2} (W-\hat W_n)^2 dxdy \le C(n)n^{-2}\le  n^{-2(1-b-\epsilon)}.
\ee

Finally, from (\ref{mean-g}) it is easy to see that 
\be\lbl{numbered}
\|\mathbf{g}-\mathbf{g}_n\|_{L^2(I)}^2=O(n^{-1})
\ee
The combination of (\ref{Gron}), (\ref{estimW}), and (\ref{numbered})
implies  (\ref{rate}).\\
$\qed$

\section{Networks on  weighted graphs}\lbl{sec.weight}
\setcounter{equation}{0}

In this section, we study a more general case of the heat equation
on convergent sequences of weighted graphs. First, we define two
graph sequences generated by a given graphon $W$ and then we
prove the convergence of the corresponding discrete problems 
to the continuum limit (\ref{sheat}).

Throughout this section, we assume that $W: I^2\to [-1,1]$ is a
symmetric measurable function. Let $\mathcal{P}_n$ denote the 
partition of $I$ into $n$ intervals, $\mathcal{P}_n=\{I_i^{(n)}, i\in [n]\}$
(see (\ref{partition})) and
$$
X_n=\left\{{1\over n}, {2\over n}, \dots, {n\over n}\right\}.
$$

The quotient of $W$ and $\mathcal{P}_n$, 
denoted $W/\mathcal{P}_n$, is the 
complete graph on $n$ nodes 
$$
W/\mathcal{P}_n =\langle [n], [n]\times [n], \bar W_n\rangle,
$$
such that weights $(\bar W_n)_{ij}$ are obtained by averaging
$W$ over the sets in $\mathcal{P}_n$
\be\lbl{average}
(\bar W_n)_{ij} =n^2 \int_{I_i \times I_j} W(x,y) dxdy.
\ee

The second sequence of weighted graphs is obtained in a  way that
is similar to the construction of $W$-random graph (cf. \cite{LovSze06})
\be\lbl{graphH}
\mathbb{H}(S_n,W)=\langle [n], [n]\times [n], \tilde W_n\rangle, \;
(\tilde W_n)_{ij}=W\left({i\over n}, {j\over n}\right).
\ee

In the remainder of this section, we prove convergence of the nonlinear 
heat equations on $W/\mathcal{P}_n$ and $\mathbb{H}(S_n,W)$ to the continuum
equation on the graphon $W$ (cf.~(\ref{sheat})). Furthermore, we show that
the former problems correspond to the discretizations of (\ref{sheat})
using the method of Galerkin and the collocation method respectively,
thus, relating the problem of justification of the thermodynamic
limit for dynamical networks to two well-known numerical schemes for
equations of mathematical physics.

We first consider the IVP for the heat equation on $W/\mathcal{P}_n$ 
\begin{eqnarray}\lbl{ode}
{d\over dt} u_i^{(n)} (t) &=& n^{-1}\sum_{j=1}^n (\bar W_n)_{ij} D\left(  u^{(n)}_j(t)-u^{(n)}_i (t) \right), \\
\lbl{ode-ic}
u_i^{(n)}(0)&=&g^{(n)}_i, \; i\in [n],
\end{eqnarray}
where
$g^{(n)}_i$ is defined in (\ref{mean-g}).

By  associating the step function $u_n(x,t)$ with $u^{(n)}(t)$ (see (\ref{step-fun})), we rewrite 
(\ref{ode}) and (\ref{ode-ic}) as 
\begin{eqnarray}\lbl{finite-int}
{\p \over \p t} u_n(x,t) &=&\int_{I} W_n(x,y) D\left(u_n(y,t) -u_n(x,t)\right)dy,\\
\lbl{finite-int-ic}
u_n(x,0) &=& g_n(x),
\end{eqnarray}
where $W_n$ and $g_n$  are the step functions
\begin{eqnarray*}
W_n(x,y) &=& \bar{W}_{ij} \;\mbox{for} \; (x,y)\in I^{(n)}_i\times I^{(n)}_j,\\
g_n(x) &=& g^{(n)}_i, \;\mbox{for} \; x\in I^{(n)}_i.
\end{eqnarray*}

\begin{rem}\lbl{rem.Galerkin}
It is instructive to note that (\ref{ode}) and 
(\ref{ode-ic}) can be viewed as the Galerkin
approximation of the IVP (\ref{sheat}) and (\ref{sheat-ic}). 
Indeed, let $H_n$ denote a finite-dimensional subspace of $L^2(I)$ 
$$
H_n=\mbox{span}\{\mathbf{\phi}_1, \mathbf{\phi}_2,\dots, \mathbf{\phi}_n\},
$$
where $\mathbf{\phi}_i=\chi_{I^{(n)}_i}$ is the characteristic function of 
$I_i^{(n)}=[(i-1)n^{-1}, in^{-1})$.

Replacing $u(x,t)$ in (\ref{sheat}) with 
$$
u_n(x,t)=\sum_{k=1}^n u^{(n)}_k(t) \phi_k (x)\in H_n
$$
and projecting the resultant equation on $H_n$, we arrive at (\ref{ode}).
\end{rem}

\begin{thm}\lbl{Galerkin-convergence} 
Let $\mathbf{u}$ and $\mathbf{u_n}$ be the solutions
of (\ref{sheat}), (\ref{sheat-ic}), and (\ref{finite-int}), (\ref{finite-int-ic}), 
respectively. Suppose $W\in L^\infty(I^2)$ and $\mathbf{g}\in L^\infty (I)$. 
Then 
\be\lbl{converge}
\| \mathbf{u} -\mathbf{u}_n\|_{C(0,T;L^2(I))} \to 0\;\mbox{as}\; n\to \infty.
\ee
\end{thm} 
\pf\; By following the lines of the proof of Theorem~\ref{coloc-rate}
(see (\ref{Gron})), 
for $\xi_n(x,t)=u_n(x,t)-u(x,t)$
we obtain
\be\lbl{Gron2}
\sup_{t\in [0,T]}\|{\boldsymbol\xi}_n(t)\|_{L^2(I)}\le \left( \|\mathbf{g}-\mathbf{g}_n\|^2_{L^2(I)}
+{C_1\|W-W_n\|_{L^2(I^2)}\over C_2}\right) \exp\{C_2 T\},
\ee
where positive constants $C_1$ and $C_2$ are  independent of $n$.
By the Lebesgue differentiation theorem,
$$
W_n\to W  \;\mbox{and} \; \mathbf{g}_n\to \mathbf{g},\;\mbox{as}\; n\to\infty,
$$
almost everywhere on $I^2$ and $I$ respectively. Thus, the statement of the theorem follows
from (\ref{Gron2}).\\
$\qed$

The heat equation on $\mathbb{H}(X_n,W)$ is analyzed in complete analogy to 
the IVP for $W/\mathcal{P}_n$. The IVP in this case remains (\ref{finite-int})
and (\ref{finite-int-ic}) modulo the definition of the step function
\be\lbl{Wntilde}
W_n(x,y) = \tilde{W}_{ij} \;\mbox{for} \; (x,y)\in I^{(n)}_i\times I^{(n)}_j.
\ee
We assume that $W(x,y)$ is a bounded symmetric measurable
function that is almost everywhere continuous on $I^2$. Then using
the observation in Lemma 2.5~\cite{BorChay11},
$$
W_n(x,y) \to W(x,y),\;\mbox{as}\; n\to\infty
$$
at every point of continuity of $W$, i.e., almost everywhere.
Thus, by the dominated convergence theorem, we have
$$
\|W-W_n\|_{L^2(I^2)}\to 0\;\mbox{as}\; n\to\infty.
$$
With this observation, the proof of Theorem~\ref{Galerkin-convergence} 
applies to the situation at hand. Thus, we have the following theorem.

\begin{thm}\lbl{collocation-convergence} 
Let $\mathbf{u}$ and $\mathbf{u_n}$ be the solutions
of (\ref{sheat}), (\ref{sheat-ic}), and (\ref{finite-int}), (\ref{Wntilde}),
(\ref{finite-int-ic}), 
respectively. Suppose $W\in L^\infty(I^2)$, $\mathbf{g}\in L^\infty (I)$,
and $W$ is continuous almost everywhere on $I^2$. 
Then 
\be\lbl{converge}
\| \mathbf{u} -\mathbf{u}_n\|_{C(0,T;L^2(I))} \to 0\;\mbox{as}\; n\to \infty.
\ee
\end{thm} 

\section{Examples} \lbl{sec.examples}
\setcounter{equation}{0}
In this section, we illustrate the results of this paper 
with several examples. First, we apply Theorem~\ref{thm.reg} to explain
the regions of continuity in the chimera states \cite{KurBat02}. Next, we discuss the 
attractors of the system of Kuramoto oscillators on multipartite
graphs.
\begin{figure}
\begin{center}
{\bf a}\;\includegraphics[height=1.8in,width=2.0in]{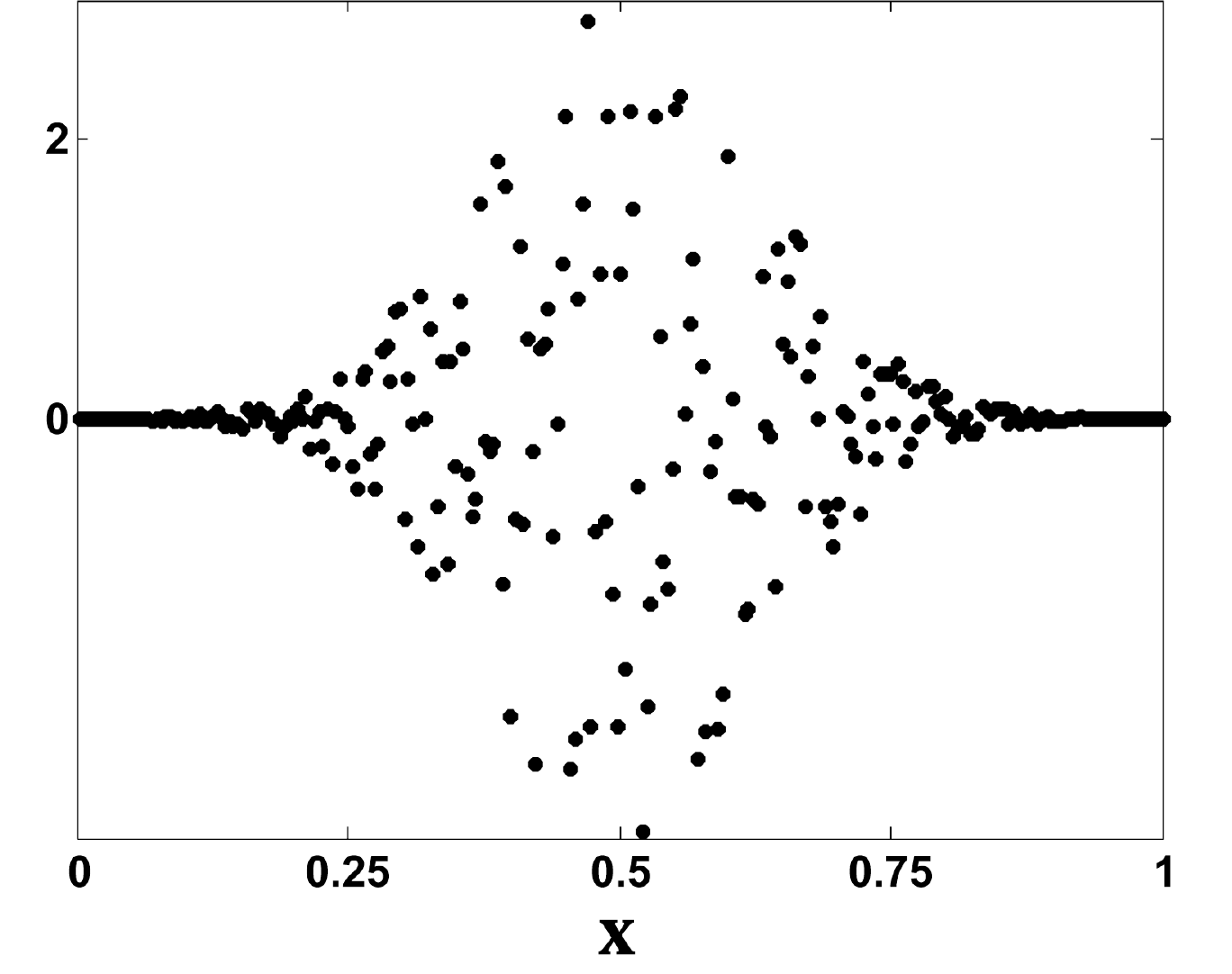}\qquad
{\bf b}\;\includegraphics[height=1.8in,width=2.0in]{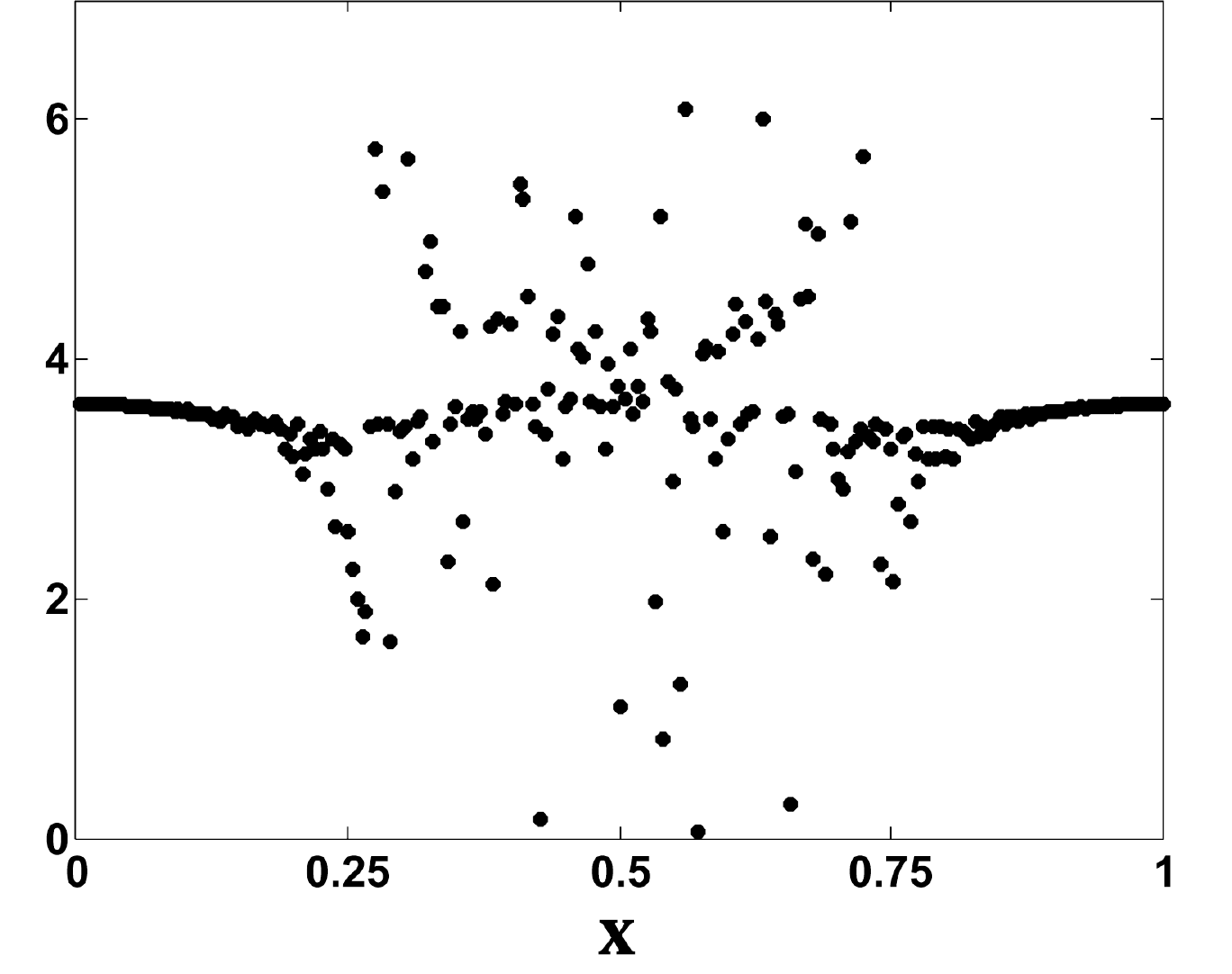}
\end{center}
\caption{ 
a) The initial conditions (\ref{chimera-ic}) for the chimera state shown in b).
b) A snapshot of the chimera state generated by (\ref{chimera}).
}
\lbl{f.2}
\end{figure}

\subsection{Regions of continuity of chimera states} \lbl{sec.chimera}
Chimera states are  persistent patterns of coexisting regions of 
spatially coherent and chaotic behaviors (see Fig.~\ref{f.2}b). 
They were discovered  by  Kuramoto and Battogtokh 
in the following  continuum limit of a system of coupled  phase oscillators \cite{KurBat02} 
\be\lbl{chimera}
{\p\over \p t} \phi(x,t) = \omega +
\int_0^1 G(x-y) \sin\left(\phi(y,t)-\phi(x,t)+\alpha\right) dy.
\ee
Function $\phi:~[0,1]\times\R^+\to \SS^1:=\R/2\pi\Z$ describes the evolution of the 
phase of  oscillator at $x\in [0,1]$. The exponential kernel $G(x)=\exp\{-\kappa|x|\}$
provides nonlocal coupling 
between oscillators. Equation (\ref{chimera}) was obtained using the phase reduction from the 
Ginzburg-Landau equation, which describes collective dynamics of nonlocally  coupled limit cycle 
oscillators (cf. \cite{KurBat02}). The sequences of discrete problems converging to (\ref{chimera})
can be obtained using one of the schemes of Section~\ref{sec.weight}.

The Kuramoto-Battogtokh model was the first example of a system featuring robust patterns that combine 
coherent and irregular dynamics. Since then chimera states were demonstrated in a variety
of computational and experimental settings \cite{LaiKar12, Show12,LarPen13}. 
The precise mathematical mechanism underlying
these patterns is the subject of ongoing research \cite{Ome13}. Here,
we focus on one aspect of the chimera states: the regions of
continuity. 
Specifically, we use Theorem~\ref{thm.reg} to explain why the synchronous 
dynamics is restricted to the two subdomains of $I$ (see Fig.~\ref{f.2}a). We show that
this possible because of the lack of the smoothening property of the
heat equation on graph limits, which is  one important 
distinction from the classical heat equation.

The numerical generation of the chimera states in (\ref{chimera}) requires a careful setup, 
which we review next.  
To trigger a chimera state one has to start with the appropriate initial
conditions, otherwise oscillators end up evolving
in phase. Abrams and Strogatz reported that they were unable to generate chimera
states in (\ref{chimera}) from smooth initial conditions \cite{AbrStr06}. Instead, one has to initialize the
system with the initial condition  that combines
the regions of coherent and incoherent spatial profiles. 
The  following initial condition was suggested by Kuramoto (cf. \cite{AbrStr06}): 
\be\lbl{chimera-ic}
\phi(x_i,0)=h(x_i) r_i, \; \mbox{where} \;  
h(x)=6\exp\left\{ -30\left(x_i -(1/2)\right)^2\right\},\; x_i=in^{-1},\; i\in [n],
\ee
and  $r_i$ are independent random variables drawn from the uniform distribution on $(-1/2, 1/2)$
(see Fig.~\ref{f.2}a).
The values of the other parameters are $\kappa=4$, $\alpha=1.457$ (cf. \cite{AbrStr06}).
 Numerical integration of
(\ref{chimera}) and (\ref{chimera-ic}) with these parameter values yields 
persistent patterns with coexisting regions of spatially coherent and chaotic dynamics.
A representative snapshot is shown in Fig.~\ref{f.2}b.

Theorem~\ref{thm.reg} explains the role of the initial conditions in generating chimera
states.
Note that function $h(x)$ in (\ref{chimera-ic}) is rapidly decaying to $0$ outside a 
neighborhood of $1/2$ . Therefore, the initial conditions
in the intervals $J_1=(0,0.2)$ and $J_2=(0.8, 1)$ near the endpoints of the interval $[0,1]$
for all practical purposes can be viewed if they were produced by discretization of a function 
that is smooth over $J_1$ and $J_2$ (see Fig.~\ref{f.2}b). For such initial conditions, 
Theorem~\ref{thm.reg} implies that the 
solution $\phi(x,t)$ will remain continuous on $J_1$ and $J_2$, because $H^1(J_{1,2})\subset C(J_{1,2})$
by the Sobolev Embedding Theorem \cite{EvaPDE}. This  explains why the spatial profile remains coherent over 
$J_1$ and $J_2$ for positive times (see Fig.~\ref{f.2}a).
Theorem~\ref{thm.reg} also implies  that it is impossible to generate chimera states starting 
from smooth initial data, because for such data the solution of the continuum limit remains
continuous over the entire domain for all $t>0$. This rules out regions of chaotic
behavior in large networks, because their solutions remain close to that of the continuous
system by Theorem~\ref{Galerkin-convergence} or Theorem~\ref{collocation-convergence}.
This explains  failed attempts to produce chimera states from smooth 
initial conditions  in \cite{AbrStr06}.

\subsection{The Kuramoto equation on multipartite graphs}\lbl{sec.half}

To illustrate our results for networks on simple graphs (see Section~\ref{sec.simple}), 
we discuss the Kuramoto equation on multipartite graphs.
The examples of this subsection illustrate another implication of the 
lack of smoothening property of the heat equation on graph limits. This time 
we show that the lack of smoothness  of the limiting graphon  may result in stable 
discontinuous patterns.

Consider the Kuramoto equation on the sequence of bipartite complete graphs
\be\lbl{halfK}
\dot u_i^{(n)} (t) ={ (-1)^\sigma\over n} \sum_{j:~(j,i)\in E(K_{n,n})} 
\sin\left( u_j^{(n)} (t)-u_i^{(n)} (t)\right),\quad i\in [2n],
\ee
where 
$$
K_{n,n}=\langle [2n], E(K_{n,n})\rangle,\;\;\mbox{and}\;\;
E(K_{n,n})=\{ (i,j)\in [n]^2:~ 1\le i\le n< j\le 2n\}.
$$
The sequence $\{K_{n,n}\}$ is  convergent with the limit shown in Fig.~\ref{f.3}a.
We consider two models for $\sigma=0$
and $\sigma=1$.
As shown below, the space homogeneous (synchronous) solution is stable 
for the $\sigma=0$ model and is unstable if $\sigma=1$.

\begin{figure}
\begin{center}
{\bf a}\includegraphics[height=1.8in,width=2.0in]{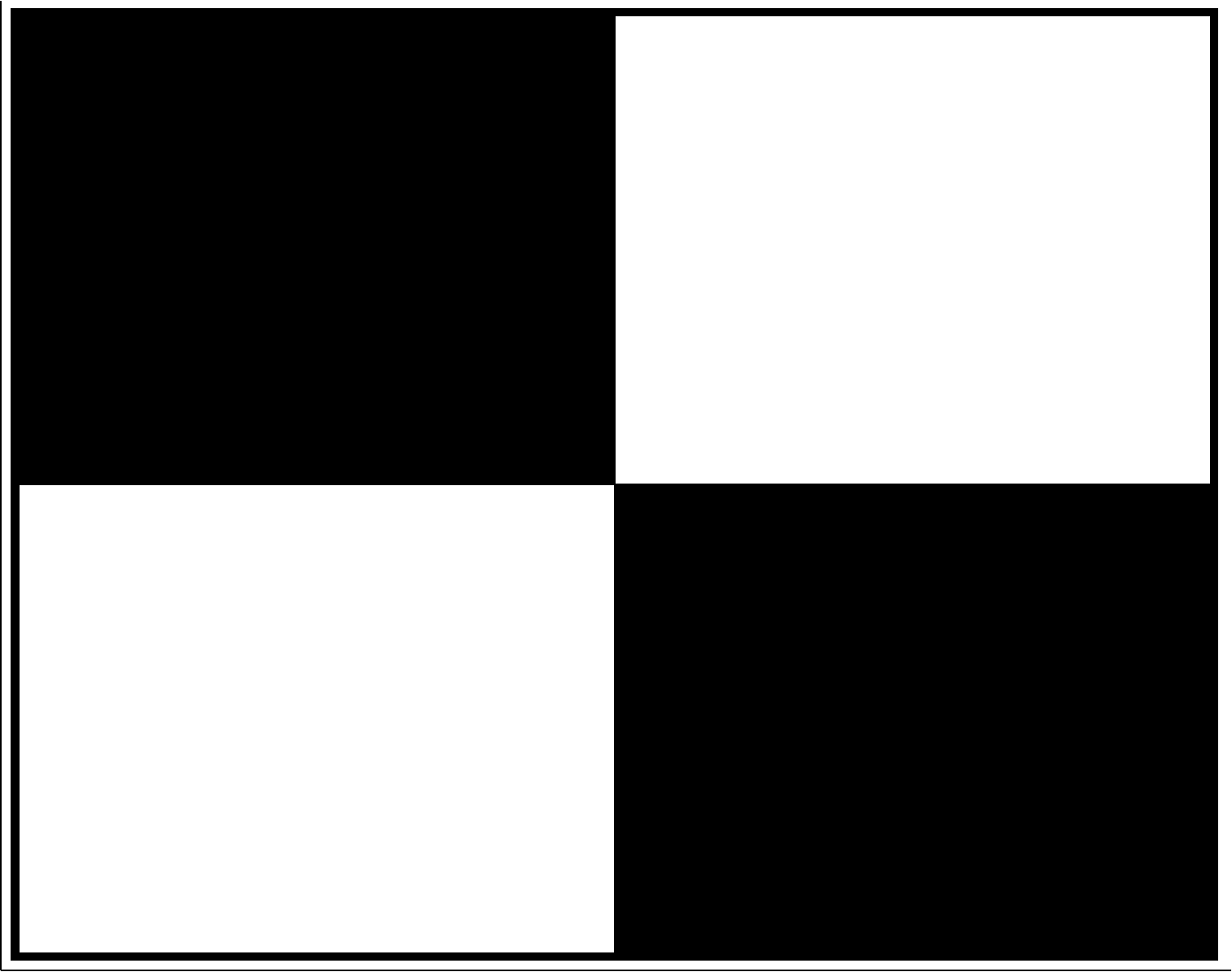}
{\bf b}\includegraphics[height=1.8in,width=2.0in]{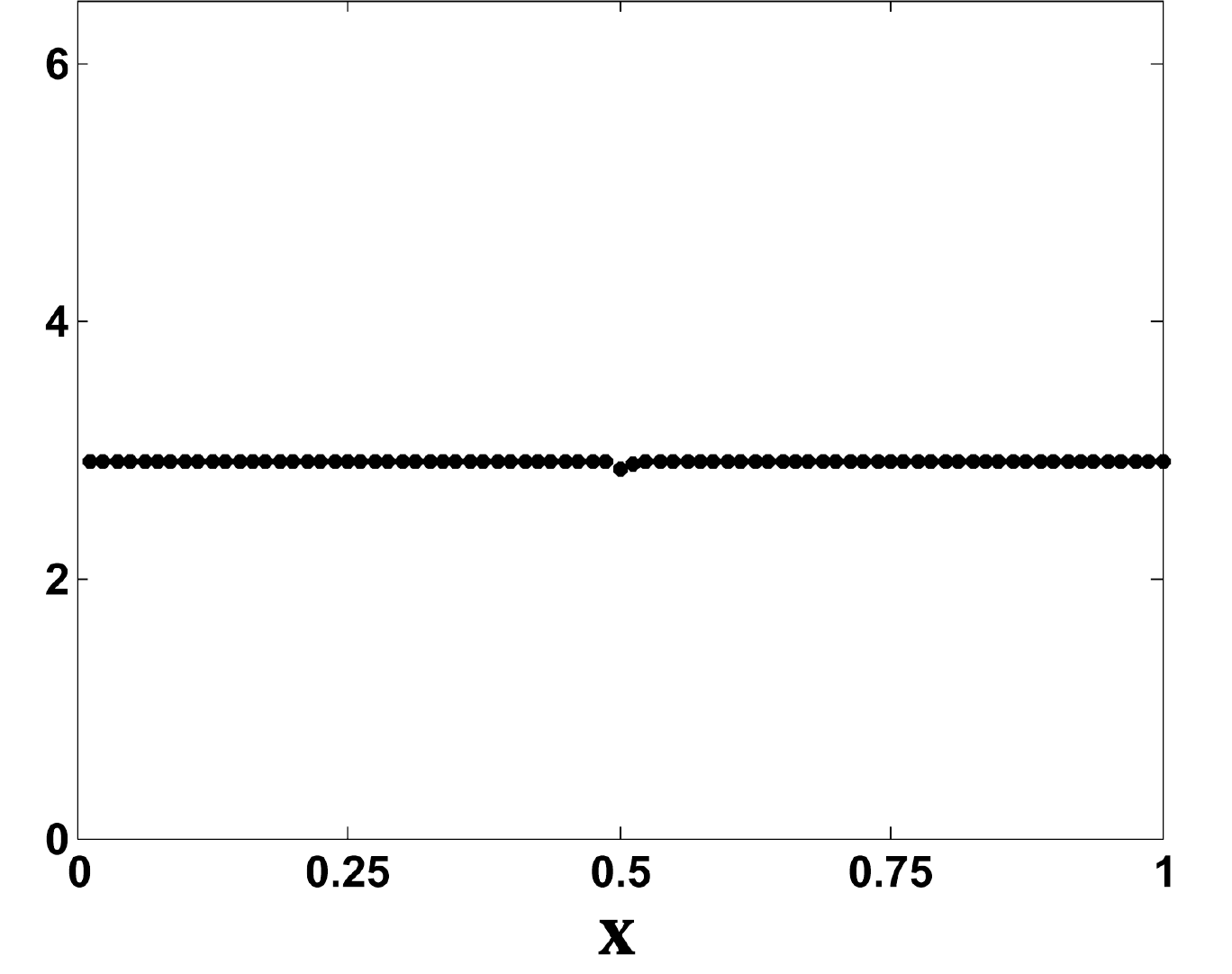}
{\bf c}\includegraphics[height=1.8in,width=2.0in]{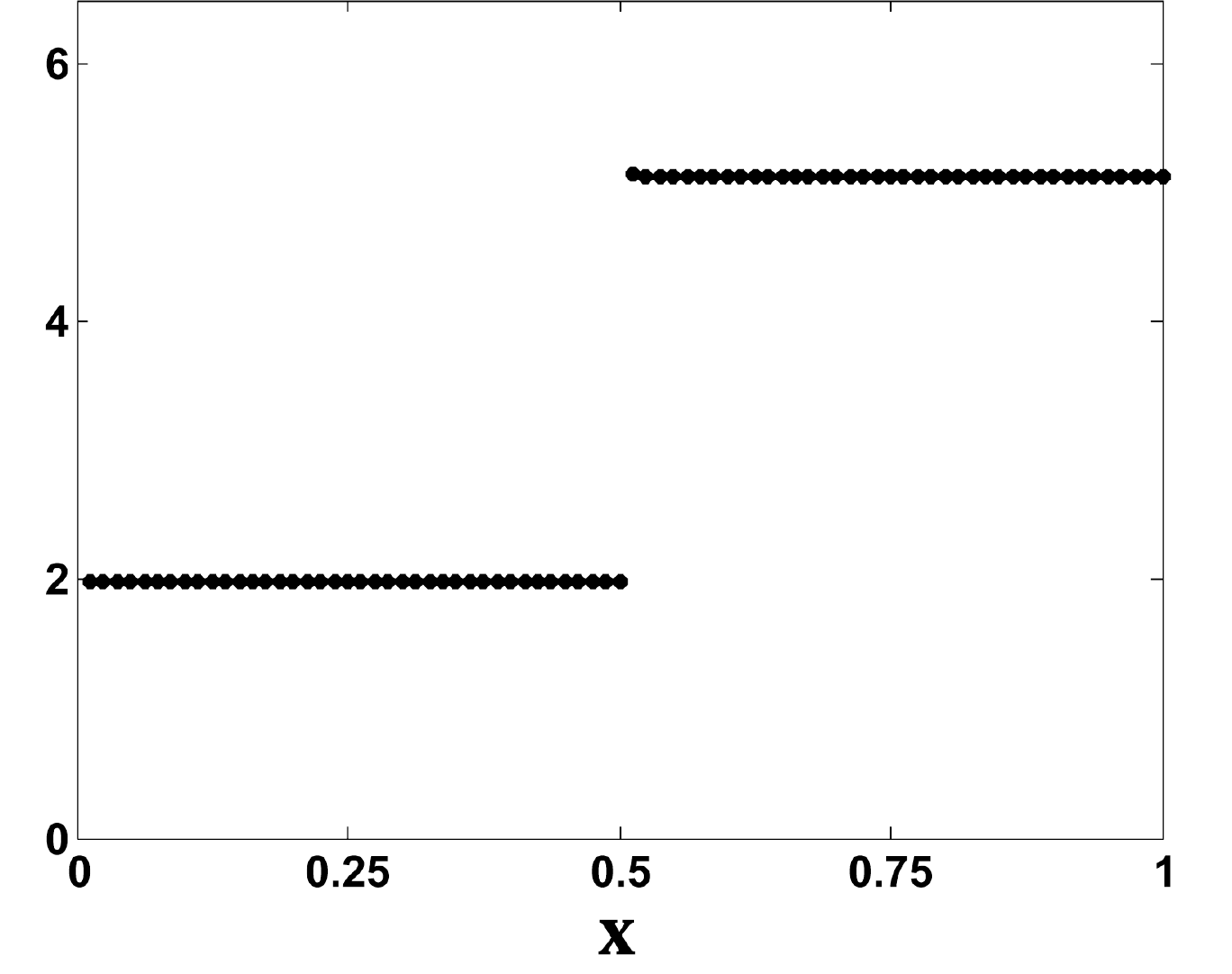}
\end{center}
\caption{ 
\textbf{a})The plot of support of $W_{K_{n,n}}$.
 \textbf{b,c}) Solutions of the IVP problem for the Kuramoto equation
on the bipartite complete graphs converge to the 
synchronous solution for $\sigma=0$ ({\bf b}) and to the step function for
$\sigma=1$ ({\bf c}).
}
\lbl{f.3}
\end{figure}

Along with (\ref{halfK}) we consider its continuum limit
\be\lbl{halfC}
{\p \over \p t} u(x,t) = (-1)^\sigma \int_I K(x,y) \sin\left( u(y,t)-u(x,t)\right) dy,
\ee
where graphon $K\in\mathcal{W}_0$ is the limit of $\{K_{n,n}\}$ (see Fig.~\ref{f.3}a).
Suppose  $u(x,0)\in C(I)$. By Theorem~\ref{thm.reg}, for any $t>0$,
$u(x,t)\in \tilde C(I)$ where
$$
\tilde C(I)=\{ u\in L^\infty(I):~ \mbox{for any open interval}~ J\subset (0,1/2)
\cup (1/2,1)~ u\left|_J\right.\in C(J)\}.
$$
Here, by $u\left|_J\right.$ we denote the restriction of $u$ to $J$.

We look for steady state solutions of (\ref{halfC}) that belong to $\tilde C(I)$.
Setting the right hand side of (\ref{halfC}) to $0$, we obtain
\begin{eqnarray}\lbl{H1}
\int_{1/2}^1 \sin\left( u(y,t)-u(x,t)\right) dy=0, & x\in (0,1/2),\\
\lbl{H2}
\int_0^{1/2} \sin\left( u(y,t)-u(x,t)\right) dy=0, & x\in (1/2, 1).
\end{eqnarray}
From (\ref{H1}) and (\ref{H2}), we find that the only piecewise constant steady state solutions
from $\tilde C(I)$ are the space homogeneous 
function
$$
u^{h}(x)=c,\;\mbox{for}\; x\in[0,1],
$$
and the step function
$$
u^s(x)=\left\{ \begin{array}{cc} c_1, & x\in [0,1/2),\\
                                               c_2, & x\in [1/2,1],
\end{array}
             \right.
$$
where constants $c,c_1,c_2\in \SS^1$ and $|c_2-c_1|=\pi$. 

Next, we turn to the discrete model (\ref{halfK}). The discrete counterparts of 
$u^{s}(x,t)$ and $u^{h}(x,t)$ are
$$
u^s=c\one_{2n}\in \R^{2n} \;\mbox{and}\; 
u^h=(c_1\one_{n}^\t, c_2\one_{n}^\t)\in \R^{2n},
$$
where $\one_{n}=(1,1,\dots,1)^\t\in\R^n$.
 
The linearization of (\ref{halfK}) about $u=u^h$ yields
\be\lbl{lin-half}
\dot \xi = { (-1)^{\sigma+1}\over n} \mathbf{L} \xi.
\ee
Matrix $\mathbf{L}$ is the Laplacian of $K_{n,n}$
\be\lbl{matrixL}
\mathbf{L}=\begin{pmatrix} nI_n & -J_n\\ -J_n & nI_n \end{pmatrix}, \;
\ee
where $I_n$ is the $n\times n$ identity matrix and $J_n=\one_n \one_n^\t$.
As a graph Laplacian of an undirected connected graph, $\mathbf{L}$ is a symmetric
positive  semi-definite matrix with a simple eigenvalue $0$ \cite{Fied73}. 
Thus, the space homogeneous solution $u^h$ is stable
for $\sigma=0$ and is unstable when $\sigma=1$.\footnote{The simple 
zero eigenvalue in the spectrum of the linearized problem reflects 
the translational invariance of (\ref{halfK}), which does not affect the
stability.}
The linearization of (\ref{halfK}) about $u^s$ yields
$$
\dot \xi ={(-1)^{\sigma+1}\over n}\mathbf{L}\xi,
$$
which, up to a sign, coincides with (\ref{lin-half}).
Thus, $u^s$ is unstable if $\sigma=0$ and is stable for $\sigma=1$.

The discrete model (\ref{halfK}) has many other piecewise constant 
steady state solutions besides $u^h$ and $u^s$. But the latter are the only two
that approximate functions in $\tilde C(I)$ and, therefore, only these solutions
can be attractors of the discrete system for large $n$ (cf. Theorem~\ref{thm.reg}).
This  is consistent with the numerical simulations shown in Fig.~\ref{f.3}b,c.
Numerical experiments show that the synchronous state is the attractor for 
the Kuramoto model with $\sigma=0$, while the step function is the attractor 
for the model with $\sigma=1$ (see Fig.~\ref{f.3}b,c).

\begin{rem}\lbl{rem.half}
The  Kuramoto model on the family of half-graphs (cf.~Example~\ref{ex.half}) 
also exhibits exhibits stable step-like patterns, whose  
analysis follows the lines  of that for the complete bipartite graphs.
\end{rem}
\begin{figure}
\begin{center}
{\bf a}\includegraphics[height=1.8in,width=2.0in]{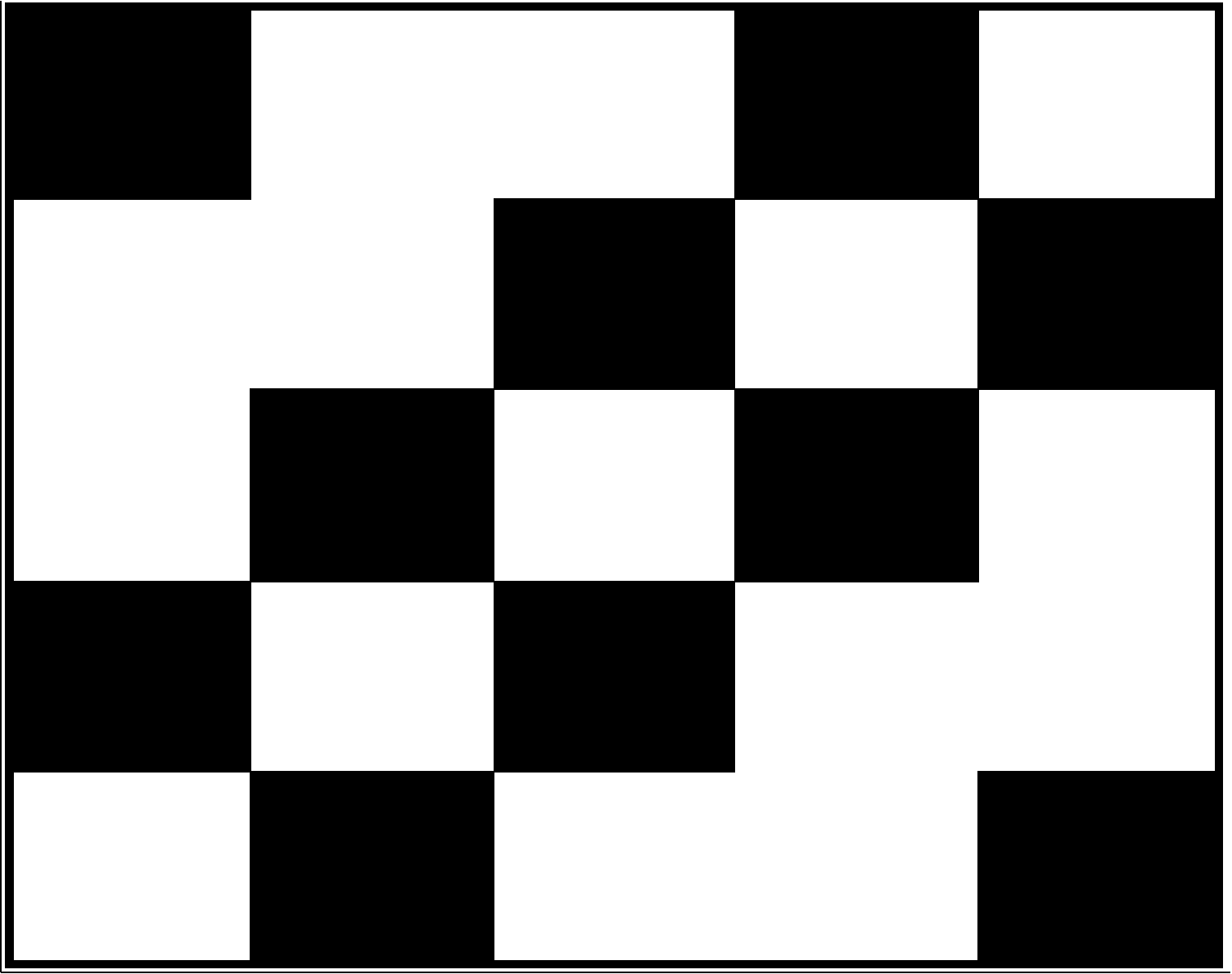}\qquad
{\bf b}\includegraphics[height=1.8in,width=2.0in]{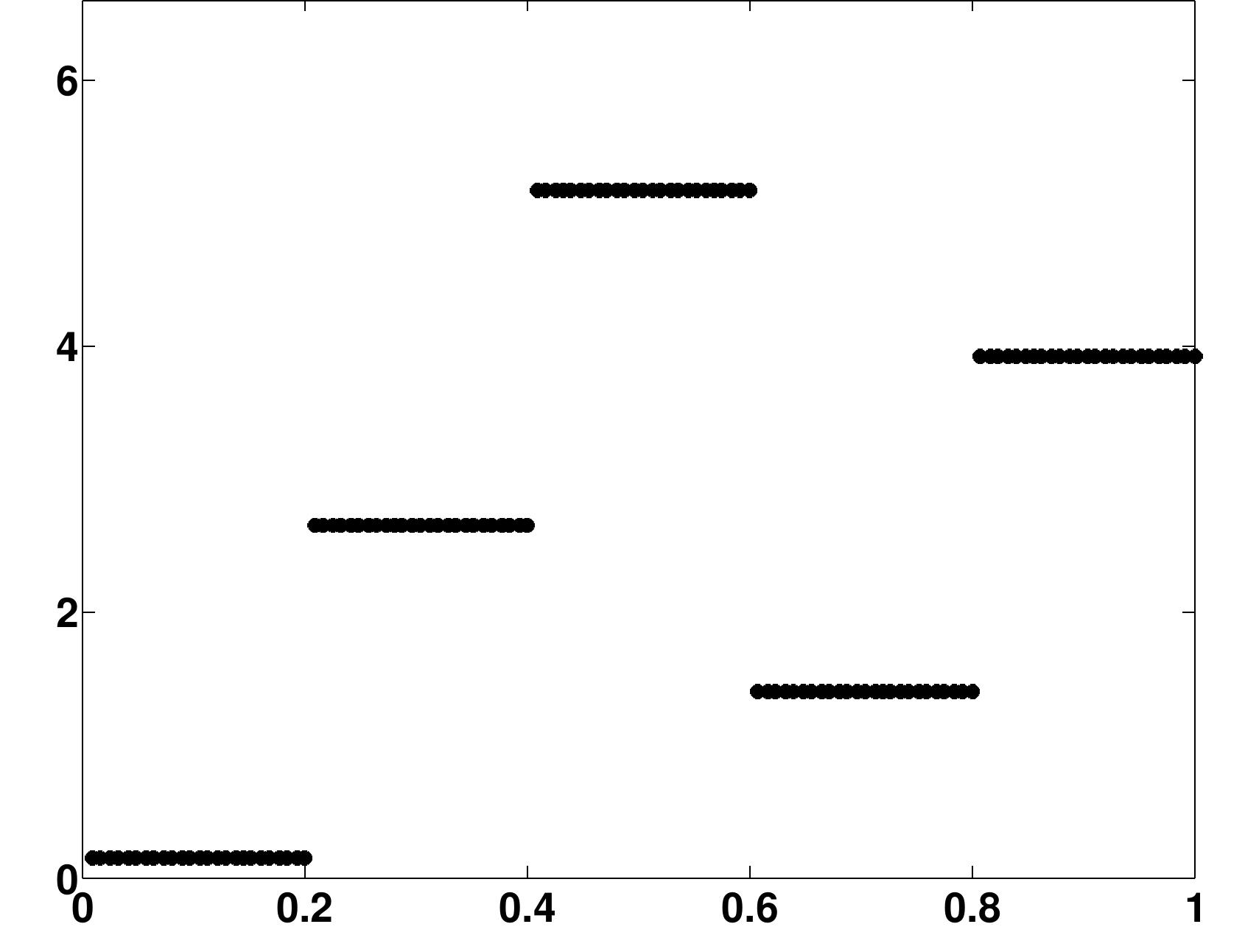}
\end{center}
\caption{ 
{\bf a}) The block structure of $A(G_{nm})$.  {\bf b}) A stable multistep pattern
generated by the Kuramoto model on a multipartite graph.   
}
\lbl{f.last}
\end{figure}
In conclusion, we briefly discuss how the Kuramoto model on $\{K_{n,n}\}$ can be generalized
to produce stable patterns with arbitrary number of steps. 
To this end, let $C_n=\langle V(C_n), E(G_n)\rangle$ be an $n$-cycle, i.e.,
$V(C_n)=[n]$ and $E(C_n)=\{(i,j)\in [n]^2:~ \mbox{dist}(i,j)=1\}$.
Recall  $\mbox{dist}(i,j):=\min\{ |i-j|, n-|i-j|\}$. 
The adjacency matrix of $C_n$ is given by 
\be\lbl{A(C)}
A(C_n)=\begin{pmatrix} 0 & 1&  0 &  \dots & 0&1\\ 1 &0 &  1 &  \dots
&0  &0\\ & & & \dots &  &\\ 1 & 0&  0 & \dots & 1& 0 \end{pmatrix}.
\ee

Let $K_m$ denote the complete graph on $m$ nodes. Define graph
$C_{n,m}=C_n\otimes K_m$
on $nm$ nodes by replacing each node of $C_n$ with a copy of the complete graph
$K_m$. The adjacency matrix of the resultant graph
is the Kronecker product of $A(C_n)$ and $A(K_m)$
$$
A(C_{n,m})=A(C_n)\otimes A(K_m).
$$
The block structure of $A(C_{n,m})$ is shown in Fig.~\ref{f.last}a. 

The Kuramoto model (\ref{halfK}) with $K_{n,n}$ replaced by $C_{n,m}$
generates stable patterns with $n$ steps like those shown in
Fig.~\ref{f.last}b.  In computational  neuroscience, such patterns
have been sought in the context of modeling memory. 
The stability analysis of these multistep patterns, which can be done in analogy
to the analysis in this subsection, will be
presented elswhere.


\section{Conclusion}\lbl{sec.conclusion}
\setcounter{section}{0}
The heat equation is a fundamental equation of mathematical
physics. On Euclidean domains, the heat operator is used to model
phenomena involving diffusion, propagation, and pattern formation
in diverse problems of physics and biology. On Riemannian manifolds,  the 
heat equation has been a powerful tool for studying the topology of the underlying
manifold \cite{Rosenberg-Laplacian}.  Its discrete counterpart, the heat equation
on graphs plays an important role in the spectral graph theory \cite{Chung-Spectral}.

Motivated by the dynamics  large networks, in this paper
we have studied the nonlinear heat equation on dense graphs. 
We identified two classes of convergent graph sequences, for which the dynamics
of large coupled networks is approximated by the heat equation on the graph 
limit. The latter is a nonlinear evolution equation with an integral operator
that describes nonlocal spatial interactions. The nonlocal heat equation differs
from its partial differential equation counterpart in several respects. First,
the IVP for the heat equation on a graph limit is well-posed in both forward
and backward time. Second, the solutions of the IVPs for the nonlocal heat equation  
lack the smoothening property, i.e., the spatial regularity of solutions for positive times
is determined by the initial data and the regularity of the graph limit. In particular,
the heat equation on a graph limit can have attractors that are  piecewise continuous 
in space (see Subsection~\ref{sec.half}), or  combine regions with qualitatively 
distict dynamics like in chimera states (see Subsection~\ref{sec.chimera}).

Our analysis highlights the properties of the convergent graph sequences that
are necessary for supporting the continuum limit for coupled dynamical systems.
Note that for convergent sequences of simple graphs analyzed in Section~\ref{sec.simple},
we require that the graph limit is a $\{0,1\}$-valued graphon. For such sequences, 
we are able to represent the discrete problems using the step functions $\{\hat W_{G_n}\}$ 
(cf.~(\ref{hat-W})), which are convergent in the $L^1$-norm. This construction does
not work for an arbitrary sequence of simple graphs. For instance,  a sequence  
of Paley graphs converges to the constant graphon equal to $1/2$,
$\mbox{Const}~(1/2)$ \cite{BorChay08}. 
However, the corresponding
continuum limit (\ref{sheat}) does not approximate the dynamics of the discrete problems.
On the other hand, the analysis in \cite{Med13a} shows that the heat equation on 
the sequence of the Erd\H{o}s-R'{e}nyi graphs  (which is also a sequence of
simple albeit random graphs converging to $\mbox{Const}(1/2)$) has a well-defined
continuum limit. In contrast to the present work, the analysis of the continuum limit in 
\cite{Med13a} does not rely on the $L^1$-norm for graphons, but effectively uses
the cut-norm.

Our results for networks on convergent sequences of simple graphs also reveal 
what properties of  graphs affect the accuracy of the continuum limit.
Specficially, the rate of convergence estimate in Theorem~\ref{coloc-rate} shows that 
the accuracy of approximation of the solutions of the discrete problems by their
continuous counterparts depends on the regularity of the boundary of support
of the graph limit. In particular, the convergence may slow down
significantly if the Hausdorff dimension of the boundary is close to $2$. It is interesting to compare this
result with the rate of convergence estimate for random networks in \cite{Med13a}.
For random networks, the rate is determined by the Central Limit Theorem and 
is independent of the regularity of the underlying graphon.

The theory of graph limits provides a useful set of tools for studying dynamics
of large networks \cite{LovGraphLim12}. On  one hand, known graph limits for various 
convergent sequences like that of half graphs or Erd\H{o}s-R\'{e}nyi graphs 
suggest  continuum limits for the corresponding networks.
On the other hand, this rich theory offers many useful ideas and analytical
results that can be applied to the analysis dynamical networks. In this paper, we analyzed
two families of networks on convergent sequences of deterministic graphs.
In \cite{Med13a, Med13b} a similar approach is used to study networks on convergent 
sequences of random graphs. Therefore, the results of this paper and in 
\cite{Med13a} justify the continuum limit for a broad class of networks.

\vskip 0.2cm
\noindent
{\bf Acknowledgements.} The  author thanks A.~Grinshpan and
D.~Kaliuzhnyi-Verbovetskyi for useful discussions and
valuable comments on the manuscript. This work was supported in part by 
the NSF grant DMS 1109367.

 \vfill\newpage
 \bibliographystyle{amsplain}
\providecommand{\bysame}{\leavevmode\hbox to3em{\hrulefill}\thinspace}
\providecommand{\MR}{\relax\ifhmode\unskip\space\fi MR }
\providecommand{\MRhref}[2]{%
  \href{http://www.ams.org/mathscinet-getitem?mr=#1}{#2}
}
\providecommand{\href}[2]{#2}

\end{document}